\documentclass{aastex63}

\usepackage{CJK}

\received{June 1, 2019}
\revised{January 10, 2019}
\accepted{\today}
\submitjournal{ApJ}

\shorttitle{Superflares in GWAC triggers}
\shortauthors{Li et al.}

\begin{document}

\begin{CJK*}{UTF8}{gbsn}

\title{The white-light superflares from cool stars in GWAC triggers}

\correspondingauthor{Guang-Wei Li}
\email{lgw@bao.ac.cn}

\author[0000-0001-7515-6307]{Guang-Wei Li (李广伟)}
\affiliation{Key laboratory of Space Astronomy and Technology, National Astronomical Observatories, Chinese Academy of Sciences, Beijing 100101, China}

\author[0000-0003-3603-1901]{Liang Wang (王靓)} 
\affiliation{Nanjing Institute of Astronomical Optics \& Technology, Chinese Academy of Sciences, Nanjing 210042, China}
\affiliation{CAS Key Laboratory of Astronomical Optics \& Technology, Nanjing Institute of Astronomical Optics \& Technology, Nanjing 210042, China }
\affiliation{University of Chinese Academy of Sciences, Beijing 100049, China}

\author[0000-0002-4554-5579]{Hai-Long Yuan (袁海龙)}
\affiliation{Key Laboratory of Optical Astronomy, National Astronomical Observatories, Chinese Academy of Sciences, Beijing 100101, China}

\author[0000-0002-9422-3437]{Li-Ping Xin (辛立平)}
\affiliation{Key laboratory of Space Astronomy and Technology, National Astronomical Observatories, Chinese Academy of Sciences, Beijing 100101, China}

\author{Jing Wang (王竞)}
\affiliation{Guangxi Key Laboratory for Relativistic Astrophysics, School of Physical Science and Technology, Guangxi University, Nanning 530004, China}
\author{Chao Wu (吴潮)}
\affiliation{Key laboratory of Space Astronomy and Technology, National Astronomical Observatories, Chinese Academy of Sciences, Beijing 100101, China}


\author{Hua-Li Li (黎华丽)}
\affiliation{Key laboratory of Space Astronomy and Technology, National Astronomical Observatories, Chinese Academy of Sciences, Beijing 100101, China}

\author{Hasitieer Haerken (哈斯铁尔·哈尔肯)}
\affiliation{ School of Artificial Intelligence of Beijing Normal University, No.19, Xinjiekouwai St, Haidian District, Beijing, 100875, China}
\affiliation{Key laboratory of Space Astronomy and Technology, National Astronomical Observatories, Chinese Academy of Sciences, Beijing 100101, China}

\author{Wei-Hua Wang (王伟华)} 
\affiliation{Changzhou Institute of Technology, Changzhou, China}

\author{Hong-Bo Cai (蔡洪波)}
\affiliation{Key laboratory of Space Astronomy and Technology, National Astronomical Observatories, Chinese Academy of Sciences, Beijing 100101, China}

\author{Xu-Hui Han (韩旭辉)}
\affiliation{Key laboratory of Space Astronomy and Technology, National Astronomical Observatories, Chinese Academy of Sciences, Beijing 100101, China}

\author{Yang Xu (徐洋)}
\affiliation{Key laboratory of Space Astronomy and Technology, National Astronomical Observatories, Chinese Academy of Sciences, Beijing 100101, China}

\author{Lei Huang (黄垒)}
\affiliation{Key laboratory of Space Astronomy and Technology, National Astronomical Observatories, Chinese Academy of Sciences, Beijing 100101, China}

\author{Xiao-Meng Lu (卢晓猛)}
\affiliation{Key laboratory of Space Astronomy and Technology, National Astronomical Observatories, Chinese Academy of Sciences, Beijing 100101, China}

\author{Jian-Ying Bai (白建迎)}
\affiliation{Key laboratory of Space Astronomy and Technology, National Astronomical Observatories, Chinese Academy of Sciences, Beijing 100101, China}

\author{Xiang-Yu Wang (王祥玉)}
\affiliation{School of Astronomy and Space Science, Nanjing University, Nanjing 210093, China}
\affiliation{Key Laboratory of Modern Astronomy and Astrophysics (Nanjing University), Ministry of Education, Nanjing 210093, China}

\author{Zi-Gao Dai (戴子高)}
\affiliation{School of Astronomy and Space Science, Nanjing University, Nanjing 210093, China}
\affiliation{Key Laboratory of Modern Astronomy and Astrophysics (Nanjing University), Ministry of Education, Nanjing 210093, China}

\author{En-Wei Liang (梁恩维)}
\affiliation{Guangxi Key Laboratory for Relativistic Astrophysics, School of Physical Science and Technology, Guangxi University, Nanning 530004, China}

\author{Jian-Yan Wei (魏建彦)}
\affiliation{Key laboratory of Space Astronomy and Technology, National Astronomical Observatories, Chinese Academy of Sciences, Beijing 100101, China}



\begin{abstract}
M-type stars are the ones that flare most frequently, but how big their maximum flare energy can reach is still unknown. We present 163 flares from 162 individual M2 through L1-type stars that triggered the GWAC, with flare energies ranging from $10^{32.2}$ to $10^{36.4}$ erg . The flare amplitudes range from $\triangle G = 0.84$ to $\sim 10$ mag. Flare energy increases with stellar surface temperature ($T_{\rm eff}$) but both $\triangle G$ and equivalent duration $\log_{10}(ED)$ seem to be independent of $T_{\rm eff}$. Combining periods detected from light curves of TESS and K2, spectra from LAMOST, SDSS and the 2.16 m Telescope, and the Gaia DR3 data, we found that these GWAC flare stars are young. For the stars that have spectra, we found that these stars are in or very near to the saturation region, and $\log_{10}(L_{\rm H\alpha}/L_{\rm bol})$ is lower for M7-L1 stars than for M2-M6 stars. We also studied the relation between GWAC flare bolometric energy $E_{\rm bol}$ and stellar hemispherical area $S$, and found that $\log_{10}E_{\rm bol}$ (in erg) increases with increasing $S$ (in cm$^2$), and the maximum flare energy  $\log_{10}E_{\rm bol, max} \geqslant \log_{10}S + 14.25$. For M7-L1 stars, there seem to be other factors limiting their maximum flare energies in addition to stellar hemispherical area.
\end{abstract}



\section{Introduction} \label{sec:intro}
Solar flares originated from the release of magnetic energy by the magnetic reconnection in the corona \citep{1991ARA&A..29..275H}. The released energy ranges from $10^{28}$ to $10^{32}$ erg \citep{2011LRSP....8....6S}. The biggest solar flare ever detected is the Carrington flare that occurred in 1859 \citep{1859MNRAS..20...13C,1859MNRAS..20...15H}, which released energy of about $4\times 10^{32}$ erg \citep{2021ARA&A..59..445H}. The Carrington Event caused a geomagnetic storm \citep{2021ARA&A..59..445H}, and interrupted telegraph services \citep{2006AdSpR..38..159B}. The aurorae can be seen even near the Equator \citep{2016AdSpR..57..257M}.
\par
Like the solar flares, stellar flares also originated from magnetic reconnections \citep{2021MNRAS.505L..79Y,2019ApJS..241...29Y}. Most stellar flare energies detected are between $10^{32} and 10^{35}$ erg and for giants, their flare energies can even be as large as $10^{38}$ erg \citep{2019ApJS..241...29Y, 2022ApJ...935..143P}. The white-light flare energies are often thought to be from blackbody radiations with a temperature of 9000 K - 14,000 K \citep{2013ApJS..207...15K}, and even as high as 42000 K \citep{2020ApJ...902..115H}, which is much higher than the photospheric temperature of M dwarfs, so flares with energies as low as  of $10^{30}$ erg can be detected in near M-type dwarfs in optical bands \citep{2017ApJ...849...36Y}.
\par
The stellar flare activity is related to rotation and also the spectral type. The faster the rotation the stronger the activity, which is the stellar activity-rotation relationship. The relationship has been found by X-ray \citep{2011ApJ...743...48W, 2016Natur.535..526W}, Ca II H \& K lines \citep{2022ApJ...929...80B,2021ApJ...910..110L}, H$\alpha$ \citep{2017ApJ...834...85N,2023RAA....23a5016L}. 
M dwarfs are more active than earlier type stars \citep{2023RAA....23h5017A} and from M0 to M6 the fraction of flare stars increases from about 10\% to over 40\% \citep{2020AJ....159...60G}. M-type stars tend to produce more frequent and powerful flares than the Sun, and the habitable zones of M dwarfs are very near the hosts \citep{2018ApJ...861L..21K}. Therefore, considering their flare energies can be thousands of times higher than the Carrington Event of the Sun and proximities of their habitable zones, the impacts of flares on habitable planets can be several magnitudes higher than those of solar flares on the Earth. The electromagnetic radiation from X-ray to radio \citep{2005ApJ...621..398O} during flares and coronal mass ejections (CMEs) can be released by powerful flares: the more powerful the flare, the more likely the CME is released \citep{2021ApJ...917L..29L}. The intense flares can release tremendous UV fluxes and CME which can destroy O$_3$ \citep{2019AsBio..19...64T}, and the ultraviolet fluxes can sterilize lives on the planet's surface \citep{2018AsBio..18.1414E}. At the same time, the atmosphere of the planet would be heated, expanded, eroded \citep{2019LNP...955.....L} and finally even disappeare \citep{2021MNRAS.500L...1A}. On the other hand, the intense flare can trigger the prebiotic chemistry and then life \citep{2018SciA....4.3302R,C8CC01499J,2020AJ....159...60G,2021NatAs...5..298C}.
\par
The biggest flares are very rare, especially the flares with energy of $\geqslant 10^{34}$ erg from M dwarfs \citep{2022ApJ...926..204H,2023MNRAS.525.1588J}, which would have important impacts on the planet's atmosphere and life \citep[e.g.][]{2023RMxAC..55...99M,2022A&A...667A..15K,2019AsBio..19...64T,2022SciA....8I9743H}.  Some biggest flares have been detected by the GWAC, EvryFlare \citep{2019ApJ...881....9H}, ASAS-SN \citep{2019ApJ...876..115S}, and NGTS \citep{2021MNRAS.504.3246J}. The amplitudes can be $\bigtriangleup R \sim 9.5$ or $\bigtriangleup V \sim 11.2$ \citep{2023MNRAS.tmp..951X}, $\bigtriangleup R \sim 9.5$ \citep{2021ApJ...909..106X}, $\bigtriangleup V > 11$ mag \citep{2016ApJ...828L..22S}, $\bigtriangleup V \sim 10$ mag  \citep{2019MNRAS.485L.136J}, and so on.

\par
As the ground instrument of Space-based multi-band astronomical Variable Objects Monitor \citep[SVOM; ][]{2016arXiv161006892W}, the Ground-based Wide Angle Cameras (GWAC) system aims to monitor afterglows of gamma-ray bursts \citep{2023NatAs...7..724X}, and thus superstellar flares were also detected by the GWAC software \citep{2021PASP..133f5001H}. In this work, we collected 163 big flares that occurred between 2017 November and 2023 March from 162 individual stars, and try to explore the mechanism behind them. These flares and properties of host stars are given in Table \ref{tab:info}.
Several of them have been carefully studied in \citet{2021ApJ...909..106X}, \citet{2023ApJ...954..142L}, \citet{2022ApJ...934...98W}, \citet{2021ApJ...916...92W}, and \citet{2023PASP..135f4201B}. In Section \ref{sec:data} we will introduce the observations and data; in Section \ref{sec:res}, we will present the properties of GWAC flares and their host stars; in Section \ref{sec:rot}, the rotation-age-activity relationship will be studied; in Section \ref{sec:maxerg} the lower limit of the maximum flare energy that a star can produce is presented; the star age and activity pattern are discussed in Section \ref{sec:dis}; and at last, the conclusion is given in Section \ref{sec:con}.

\begin{deluxetable*}{lllllllll}
		\tablecaption{GWAC flares and properties of stars \label{tab:info}}
		\tablewidth{700pt}
		\tabletypesize{\scriptsize}
		\tablehead{ \colhead{Star No.}&\colhead{GWAC Name}&\colhead{TIC/K2}&\colhead{Simbad Name}&\colhead{DR3 Name}&\colhead{RAJ2000}&\colhead{DEJ2000}&\colhead{Gmag}& \colhead{...} \\
		& & & & &\colhead{(degree)}&\colhead{(degree)}&\colhead{(mag)}&\colhead{...}
		 }
\startdata
  1&GWAC181208A&TIC 456482672&ATO J000.1128+13.6255&Gaia DR3 2767338005380427264&0.112725&13.62563&14.740&...\\
  2&GWAC220106A&TIC 432551405&2MASS J00013265+3841525&Gaia DR3 2880981530065870720&0.386113&38.69797&18.193&...\\
  3&GWAC180116A&TIC 357411008&&Gaia DR3 2769578225960672512&2.798485&15.87654&16.542&...\\
  4&GWAC181206A&TIC 405305098&2MASS J00113451+0659388&Gaia DR3 2742685889532913024&2.893805&6.99415&16.826&...\\
  5&GWAC211124A&TIC 51940383&V* BI Psc &Gaia DR3 2767679884775564928&3.176556&13.13609&15.637&...\\
 ...
		\enddata
		\tablecomments{\\
			$a$ 'EB' means eclipse binary, and 'WDMS' means white dwarf-main sequence binaries.\\
			$b$ The spectral type is from their $G_{\rm BP} - G_{\rm RP}$.\\
        (The full machine-readable form is available at \url{https://nadc.china-vo.org/res/r101350/})  
		}
\end{deluxetable*}
	
\section{Observations and Data} \label{sec:data}

\subsection{GWAC} \label{subsec:gwac}
GWAC has 36 cameras and can cover about $\sim5000$ deg$^2$ of sky \citep{2021ApJ...909..106X,2020AJ....159...35W}. There is a 4K $\times$ 4K CCD for each camera and the pixel size is about $11''.7$. The cadence of GWAC is 15 s: 10 s for exposure and 5 s for readout. There is no filter on each camera with a limit on magnitude about $G \sim 15 $ mag, but in bad weather the limit magnitude would be much brighter. For a transit that suddenly apparent on a CCD, the GWAC software \citep{2021PASP..133f5001H} triggers one of the two Guangxi-NAOC 60 cm optical telescopes (F60A and F60B) to follow up immediately to check the transit. 
\par
We obtained 163 big flares from 162 stars in the GWAC trigger database and all were confirmed by an F60. The flare amplitude $\bigtriangleup G$ ranges from 0.83 to $\sim10$ mag. Some GWAC images were not saved due to the overload of the master computer and some unknown bugs in the communication program, which is responsible for communications between more than 100 control and calculation computers for GWAC cameras. As a result, we obtained images for 147 GWAC flares and made 147 movies for them.
These GWAC light curve data, their plots, and movies are available at \url{https://nadc.china-vo.org/res/r101350/}.
\par

\begin{figure}
  \centering
  \includegraphics[scale=0.8]{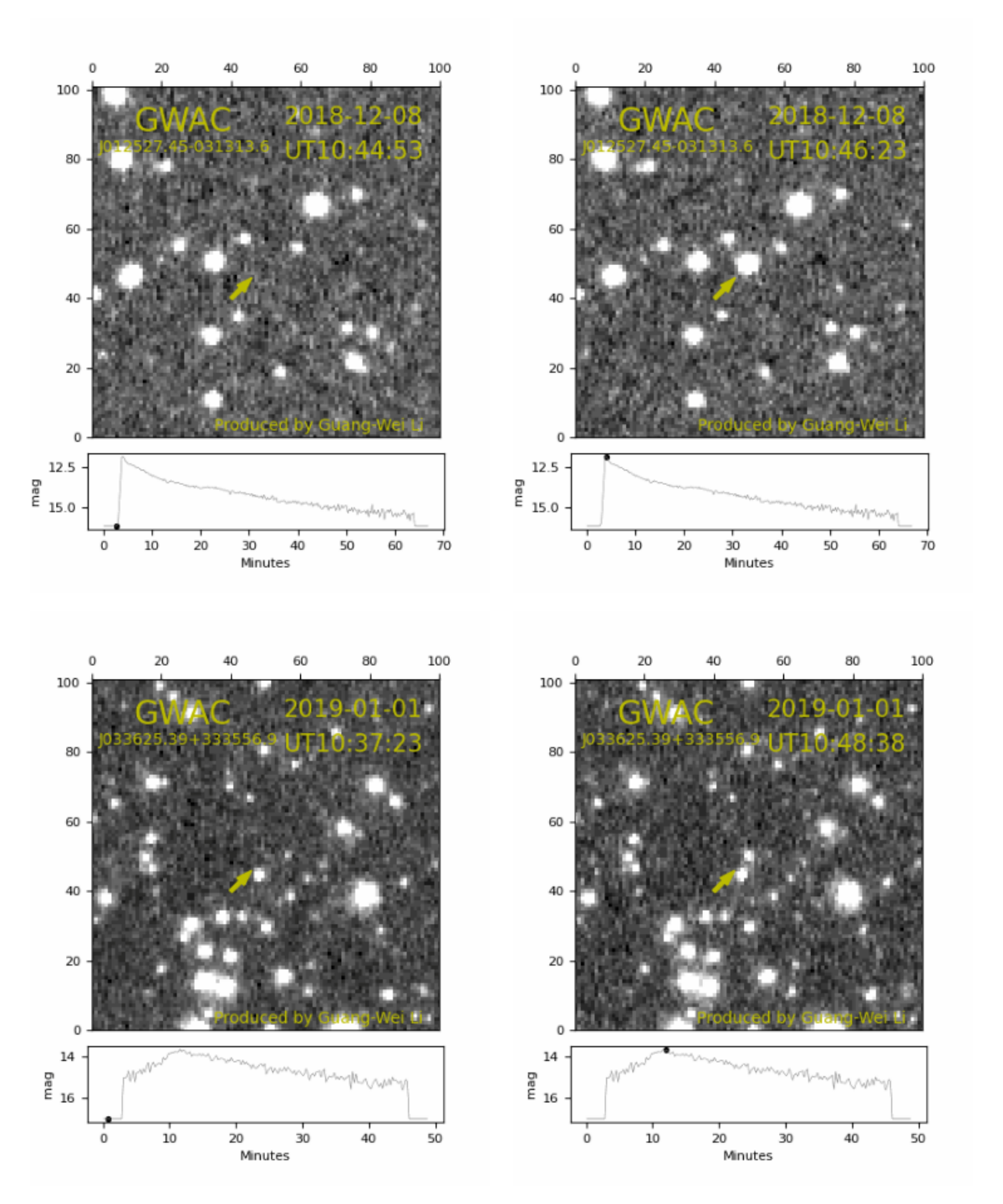}
  \caption{Two GWAC flares: GWAC181208B and GWAC190101A. The upper two panels are for GWAC181208B (from Star \#18), and the bottom two panels are for GWAC190101A (from Star \#39). The left two panels are their preflare statuses and the right two panels are the statuses at peaks. In each panel, the upper is the GWAC image, and the bottom is the GWAC light curve with time in minutes on the x-axis and flare amplitude on the y-axis. The black filled circle on the light curve indicates the flare amplitude in the image. } \label{fig:gwacflares}
\end{figure}

In Figure \ref{fig:gwacflares}, the flare GWAC181208B from Star \#18 and the flare GWAC190101A from Star \#39 are shown, and the entire flare process can been seen in their movies. The upper two panels are for GWAC181208B, and the bottom two panels are for GWAC190101A. The left two panels show their pre-flare statuses and the right two panels show flare peaks. Star \#18 and Star \#39 cannot be seen in the upper left and bottom left panels, respectively, in their preflare status, while they appeared at peaks in the upper right and bottom right panels, respectively. The flare GWAC181208B has a common flare profile: a rapid rise (lasting only about 1 minute) followed by a slow decay, with a flare amplitude of $\sim 3.3$ mag, but for the GWAC190101A,  its impulsive phase is much slower, which lasted for about 10 minutes.

\subsection{TESS and K2 Light Curves} \label{subsec:tlc}
We searched TESS light curves using the TESS-point Web Tool and found that except Star \#147 (GWAC191226A), all other stars have been observed by TESS. We searched TESS and K2 light curves by the Python package, \emph{Lightkurve} \citep{2018ascl.soft12013L}, and found available light curves for 109 stars. For the remaining 52  stars, we extracted light curves from their \emph{TargetPixelFiles}. 
For all stars but Star \#147 (GWAC191226A), we inspected their Full Frame Images (FFIs) of TESS in MAST \citep{https://doi.org/10.17909/3y7c-wa45}
and also Aladin \citep{2000A&AS..143...33B} by eye to ensure all light curves are not contaminated by nearby bright stars. For all TESS light curves of 161 stars, the bad parts of light curves were removed by hand. The K2 light curves were also obtained if available. Finally, we obtained good light curves from 276 Sectors for 124 stars. From these light curves, we tried to obtain their periods and flares using the algorithm from \citet{2023RAA....23a5016L}. In short, the algorithm tries to fit the light curve by a B-spline by iteratively removing flares, then obtains the period by the Lomb-Scargle method \citep{1982ApJ...263..835S,1976Ap&SS..39..447L} using \emph{LombScargle} in \emph{Astropy} \citep{2013A&A...558A..33A,2018AJ....156..123A}. After minus the fitted B-spline from the original light curve, flares with 3 successive points higher than 5$\sigma$ were detected. Finally, 1478 flares from 117 stars and periods for 105 stars were obtained. An example can be seen in Figure 1 in \citet{2023RAA....23a5016L}. All TESS and K2 light curves used in this work, with flares and periods detected from these light curves are available at \url{https://nadc.china-vo.org/res/r101350/}. 
\subsection{Flare Energy} \label{subsec:erg}
We calculated the equivalent duration \citep[ED;][]{1972Ap&SS..19...75G} of each flare: $$ED = \int^{t1}_{t0}\frac{f(t)-f_0}{f_0}dt$$. Here, $t_0$ and $t_1$ are the flare start and end times in second respectively, $f(t)$ is the flare flux in erg\,s$^{-1}$ at the time $t$, and $f_0$ is the stellar quiescent flux in erg\,s$^{-1}$. Then the flare energy $E= ED\times f_0$. 
\par
GWAC has no filter and we used Gaia $G$ to calibrate its photometry. The accuracy is better than 0.1 mag. We used the method in  \citet{2023RAA....23a5016L} to calculate the quiescent flux of a star, the zero point flux of $G$ and $T$ passbands are from \citet{2023RAA....23a5016L} and \citet{2015ApJ...809...77S}, respectively. The parallaxes of stars are from Gaia DR3 \citep{2022yCat.1355....0G}. Star \#9 (GWAC171207A), \#95 (GWAC210112A), \#124 (GWAC210217A), and \#104 (GWAC210202A), have no available parallaxes in Gaia DR3,  so their distances are estimated by $M_{Ks} = 1.844 + 1.116(V-J)$ \citep{2020A&A...637A..22R}, where $Ks$ and $J$ are from 2MASS \citep{2006AJ....131.1163S} and $V$ is from TIC 8.2 \citep{2022yCat.4039....0P}.
For Star \#11 (GWAC181227A), \#44 (GWAC190204A), \#49 (GWAC211203A), \#86 (GWAC210117A),  \#117 (GWAC190116B) and \#120 (GWAC190117A), their flare energies were calculated from their K2 light curves, so the method in \citet{2013ApJS..209....5S} was used, with stellar radii and surface temperatures from \citet{2016ApJS..224....2H}. Finally, a blackbody with $T = 9000$ K was used to estimate the bolometric energy that a flare released. These flares are listed in Table \ref{tab:flare}.

\begin{deluxetable*}{llll}
		\tablecaption{Flares from TESS or K2 light curves \label{tab:flare}}
		\tablewidth{700pt}
		\tabletypesize{\scriptsize}
		\tablehead{ \colhead{TIC/K2}&\colhead{BeginTime}&\colhead{EndTime}&\colhead{$\log_{10}E_{bol}$}\\
		 &\colhead{(days)}&\colhead{(days)}&\colhead{{erg}}
		 }
\startdata
TIC 156817806&2611.52955 &2611.55733 &31.78 \\
TIC 156817806&2628.24472 &2628.30028 &32.35 \\
TIC 156817806&2634.60568 &2634.68901 &32.26 \\
TIC 156817806&2635.53621 &2635.58482 &32.05 \\
TIC 318929976&1436.53726 &1436.68309 &33.38 \\
TIC 13936933&1784.47441 &1784.62024 &33.03 \\
TIC 118768009&2475.85775 &2475.86748 &33.17 \\
TIC 118768009&2495.11225 &2495.14420 &33.88 \\
TIC 118768009&2497.71940 &2497.73468 &33.63 \\
TIC 303471889&2466.06797 &2466.17215 &33.85 \\
TIC 303471889&2466.94999 &2466.99861 &33.55 \\
 ...
		\enddata
		\tablecomments{\\
			'BeginTime' and 'EndTime' are Julian days (JD 2547000 days) when a flare starts and ends.\\
        (The full machine-readable form is available at \url{https://nadc.china-vo.org/res/r101350/})  
		}
\end{deluxetable*}

\subsection{Spectra } \label{subsec:spec}
We searched low-resolution spectra ($R \sim 1800$) in LAMOST DR10 \footnote{\url{http://www.lamost.org/dr10/}} \citep{2012RAA....12..735D, 2012RAA....12.1197C} and SDSS DR17  \footnote{\url{https://www.sdss4.org/dr17/}}, and found 68 low-resolution spectra for 46 stars and 17 spectra for 12 stars, respectively. We also obtained 69 spectra for 67 stars with a resolution of 2.34\,\AA \, pixel$^{-1}$ with the instrument G5 on the 2.16 m telescope \citep{2018RAA....18..110Z}. There are five stars observed by LAMOST and also by the 2.16 m Telescope, so 154 spectra for 120 spectra were obtained. Because the resolution of spectra obtained with the 2.16 m Telescope is too low, there is no available radial velocity can be calculated from them.
\par
\begin{figure}
  \centering
  \includegraphics[scale=1]{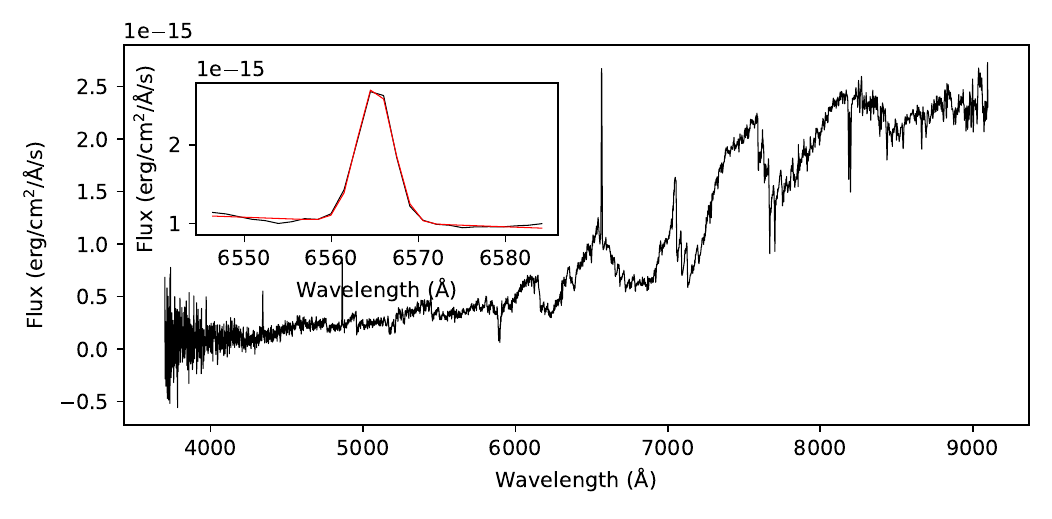}
  \caption{A LAMOST spectrum of TIC 336800108. The red line in the insert panel is the fitted H$\alpha$ emission. } \label{fig:lamostha}
\end{figure}

Except for LAMOST spectra, all other spectra are flux calibrated. Then the photometries of Pan-STARRS1 $g$ and $r$ \citep{ {2016arXiv161205560C}} were, respectively, used to calibrate LAMOST spectra for stars with $r<15$ mag and $r>15$ mag. The zero point fluxes and the filter transmission curves of Pan-STARRS1 $g$ and $r$ are obtained from the Filter Profile Service\footnote{\url{http://svo2.cab.inta- csic.es/theory/fps/}} of the Spanish Virtual Observatory. All spectra cover H$\alpha$, and the equation 
\begin{equation} \label{equ:sec}
F(\lambda) = A_0 \exp(-|(\lambda - A_1)/A_2|^{A_3}/A_3)  + A_4 \lambda + A_5
\end{equation}
 is used to fit the H$\alpha$ emission, where $\lambda$ is the wavelength in \AA, $F(\lambda)$ is the spectral flux, and $A_i, i=0,1,2,3,4,5$ is the coefficients to be fitted. Then the apparent H$\alpha$ luminosity $l_{{\rm H}\alpha} = \int_{w_1}^{w_2}A_0 \exp(-|(\lambda - A_1)/A_2|^{A_3}/A_3) d\lambda$, where $w_1 = 6544.61 $ \AA\,\,and  $w_2 = 6584.61 $ \AA. Figure \ref{fig:lamostha} shows one LAMOST spectrum of TIC 336800108 (Star \#12; GWAC181129A), with the fitted H$\alpha$ emission shown by the red curve in the insert panel. The function consists of an exponential function $A_0 \exp(-|(\lambda - A_1)/A_2|^{A_3}/A_3) $ and a linear function $A_4 \lambda + A_5$. The linear function is used to fit the spectral continuum and the exponential function is used to fit the H$\alpha$ emission. The LAMOST instrumental profile is not perfectly gaussian, and this exponential function can fit LAMOST spectral line profiles very well (see Figure \ref{fig:lamostha}). In fact, if $A_3$ is set to 2, then the exponential function would be the gaussian function.
\par
To obtain the apparent bolometric flux of a star $m_{\rm bol}$, we firstly calculated the bolometric correction of $BC_J$ using the equation in \citet{2020A&A...642A.115C}:
\begin{equation}\label{equ:cj}
BC_{J} = 0.0115 (G-J)^3 -0.132(G-J)^2 +0.735(G-J) + 0.576
\end{equation}
, where $J$ is from 2MASS and $G$ from Gaia DR3. Then $m_{\rm bol} = J+BC_J$. The apparent luminosity $l_{\rm bol}$ in erg\,cm$^{-2}$\,s$^{-1}$ of a star can be obtained by the definition of apparent bolometric magnitude in \citet{2015arXiv151006262M} by the equation
\begin{equation}\label{equ:lbo}
\log_{10}(l_{\rm bol})  = 0.4\times(-18.997351-m_{\rm bol}) +3
\end{equation}
That is,
\begin{equation}\label{equ:lboj}
\log_{10}(l_{\rm bol})  = 0.4\times(-18.997351- J-BC_J) +3
\end{equation}
Then the fraction of H$\alpha$ luminosity to bolometric luminosity $L_{\rm H\alpha}/L_{\rm bol} = l_{\rm H\alpha}/l_{\rm bol}$. Thus, $\log_{10}(L_{\rm H\alpha}/L_{\rm bol} ) = \log_{10}(l_{\rm H\alpha})-\log_{10}(l_{\rm bol}) $. $l_{\rm H\alpha}$, $l_{\rm bol}$, $\log_{10}(L_{\rm H\alpha}/L_{\rm bol})$ and radial velocities obtained from spectra with errors are given in Table \ref{tab:spec}.  
\par
For stars with available spectra, there are 3 stars: Star \#94 (GWAC180218A), \#107 (GWAC200321A), and  \#125 (GWAC200317A), have no available H$\alpha$ emission data because of low signal-to-noise fluxes around H$\alpha$ emissions.

\begin{deluxetable*}{llllllllll}
		\tablecaption{$l_{\rm H\alpha}$, $l_{\rm bol}$, $\log_{10}(L_{\rm H\alpha}/L_{\rm bol})$ and radial velocities obtained from spectra  \label{tab:spec}}
		\tablewidth{700pt}
		\tabletypesize{\scriptsize}
		\tablehead{ \colhead{Star No.}&\colhead{Telescope}&\colhead{$\log_{10}(l_{\rm bol})$}&\colhead{$\log_{10}(l_{\rm bol})$\_err}&\colhead{$\log_{10}(l_{\rm H\alpha})$}&\colhead{$\log_{10}(l_{\rm H\alpha})$\_err}&\colhead{$\log_{10}(L_{\rm H\alpha}/L_{\rm bol})$}&\colhead{$\log_{10}(L_{\rm H\alpha}/L_{\rm bol})$\_err}&\colhead{rv}&\colhead{rv\_err}\\
		 &&\colhead{[erg\,cm$^{-2}$\,s$^{-1}$]}&\colhead{[erg\,cm$^{-2}$\,s$^{-1}$]}&\colhead{[erg\,cm$^{-2}$\,s$^{-1}$]}&\colhead{[erg\,cm$^{-2}$\,s$^{-1}$]}&&&\colhead{km\,s$^{-1}$}&\colhead{km\,s$^{-1}$}
		 }
\startdata
1&LAMOST&-10.10&0.0256&-13.76&0.0065&-3.65&0.0321&11.9&3.2\\
1&LAMOST&-10.10&0.0256&-13.72&0.0067&-3.62&0.0323&15.2&3.7\\
3&216&-10.58&0.0348&-14.07&0.0052&-3.49&0.0400&&\\
4&216&-10.81&0.0252&-14.22&0.0108&-3.41&0.0360&&\\
6&216&-10.46&0.0260&-14.30&0.0122&-3.83&0.0382&&\\
 ...
		\enddata
		\tablecomments{      The full machine-readable form is available at \url{https://nadc.china-vo.org/res/r101350/}
		}
\end{deluxetable*}

\section{Results} \label{sec:res}
\subsection{GWAC flares} \label{subsec:sample}
\begin{figure}
  \centering
  \includegraphics[scale=0.8]{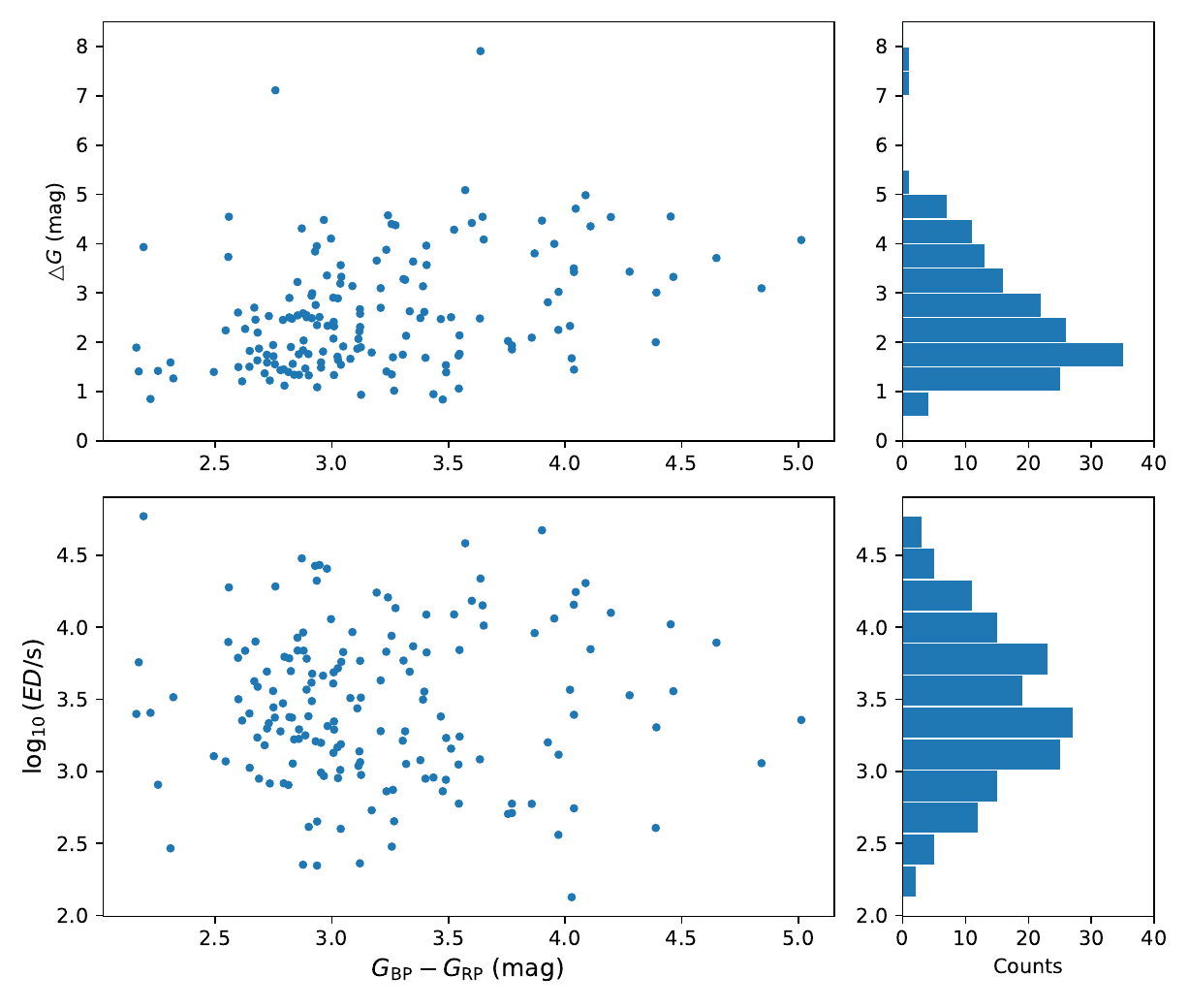}
  \caption{The distributions of $\bigtriangleup G$ and $\log_{10}(ED)$ of GWAC flares. } \label{fig:amp}
\end{figure}

The $\bigtriangleup G$ vs. $G_{\rm BP}-G_{\rm RP}$ and $\log_{10}(ED)$ vs. $G_{\rm BP}-G_{\rm RP}$ diagrams are shown in the upper and bottom panels, respectively, in Figure \ref{fig:amp}. Star \#19 (GWAC181229A) has the biggest $\bigtriangleup G \sim 10$ mag, but it is too faint to have available Gaia photometry and had been well studied in \citet{2021ApJ...909..106X}, so it is not shown in this figure.  From the upper two panels one can see that most $\triangle G$ are between 1 and 2.5 mag, and $\bigtriangleup G$ seems to be independent of $G_{\rm BP} - G_{\rm RP}$. From two lower panels, $\log_{10}(ED)$ also seems to be independent of $G_{\rm BP}-G_{\rm RP}$, or the stellar surface effective temperature.
\par
\begin{figure}
  \centering
  \includegraphics[scale=0.65]{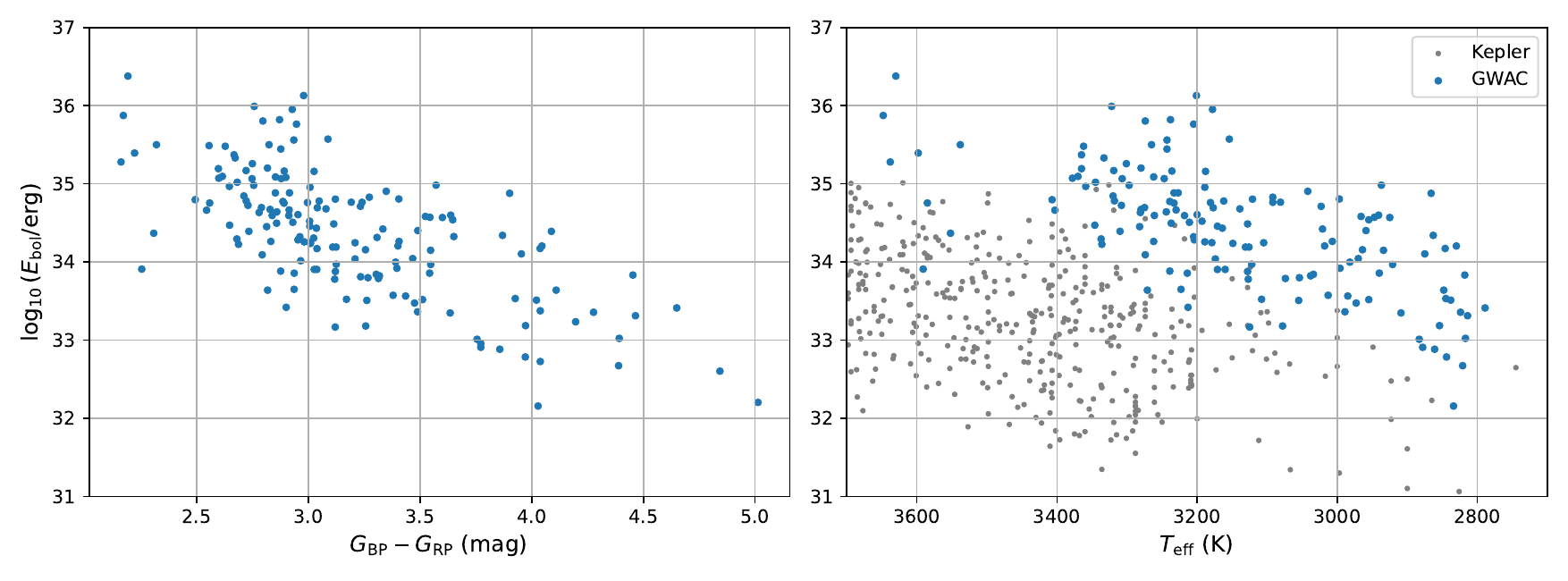}
  \caption{The bolometric energies of GWAC flares. The blue circles are energies from GWAC flares, and the gray circles are the maximum flare energies of individual stars from \citet{2019ApJS..241...29Y}. 
  } \label{fig:erg}
\end{figure}

The bolometric energies of GWAC flares are shown by blue circles in Figure \ref{fig:erg}. Stars that have available stellar surface effective temperatures in TIC 8.2 \citep{2022yCat.4039....0P} are shown in the right panel. From Figure \ref{fig:erg} we can see that the flare energy ranges from $10^{32.2}$ to $10^{36.4}$ erg, and decreases with  $G_{\rm BP}-G_{\rm RP}$ (increases with the stellar effective temperature). 
 \par
For comparison, the maximum flare energy of each star in \citet{2019ApJS..241...29Y} from Kepler DR25 is shown in the right panel in gray in Figure \ref{fig:erg}, from which we can see that the flare energies of GWAC triggers are higher than the maximum ones recorded by Kepler. \citet{2019ApJS..241...29Y} estimated flare energies from the Kepler band, while we estimated flare energies from the $G$ band. However, both assumed that bolometric flare energies are from a blackbody with a temperature of 9000 K. There are 402 Kepler stars with effective temperatures in range of 2700-3700 K for comparison, and each star was monitored by Kepler for 4 yr, which implies that GWAC superflares are at most once every 4 yr in Kepler cool stars. We also compared GWAC superflares with those detected by NGTS shown in Figure 4 in \citet{2023MNRAS.525.1588J},  and found that GWAC superflares have similar energies as those in \citet{2023MNRAS.525.1588J}, and significantly higher than those in \citet{2021MNRAS.504.3246J}, which implies that these superflares may be from the same category of top energetic flares.

\subsection{Spectral Type} \label{subsec:spt}
\begin{figure}
  \centering
  \includegraphics{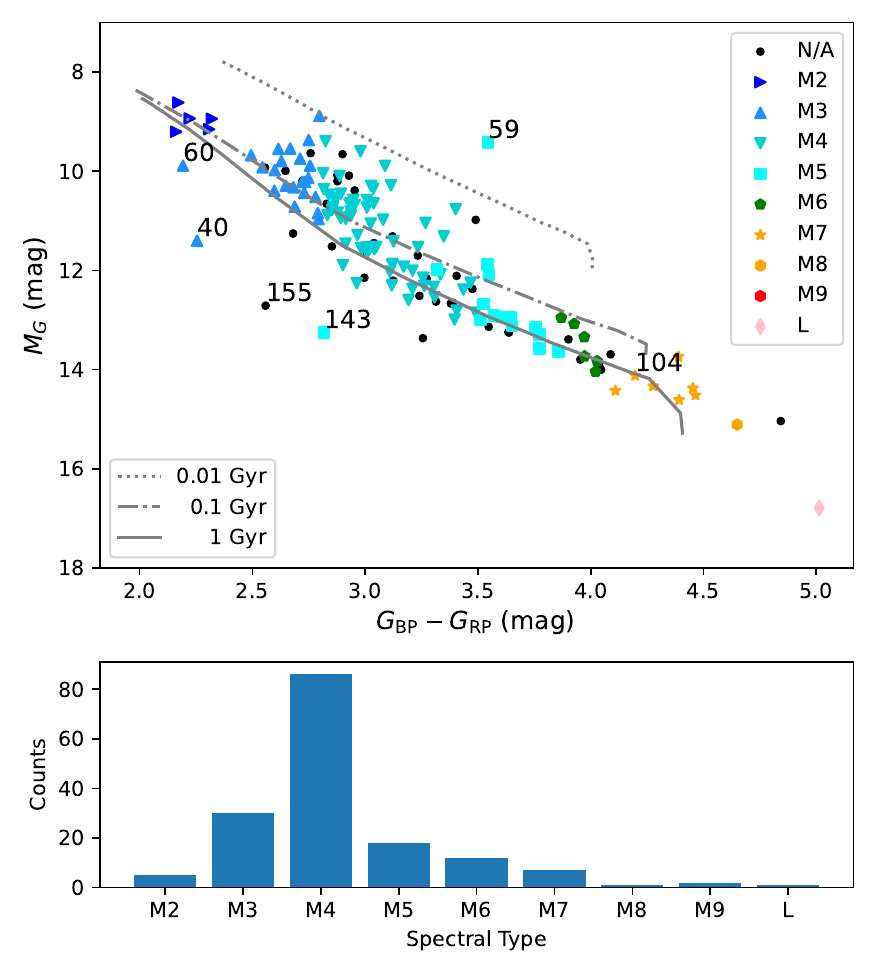}
  \caption{The H-R diagram and spectra type distribution of GWAC flare stars.} \label{fig:cmd}
\end{figure}

The spectral types were assigned by comparing the standard spectra in \citet{1991ApJS...77..417K}. Using the parallaxes and photometries from Gaia DR3 \citep{2022yCat.1355....0G}, the H-R diagram is given in Figure \ref{fig:cmd}. 
\par
\begin{figure}
  \centering
  \includegraphics[scale=1]{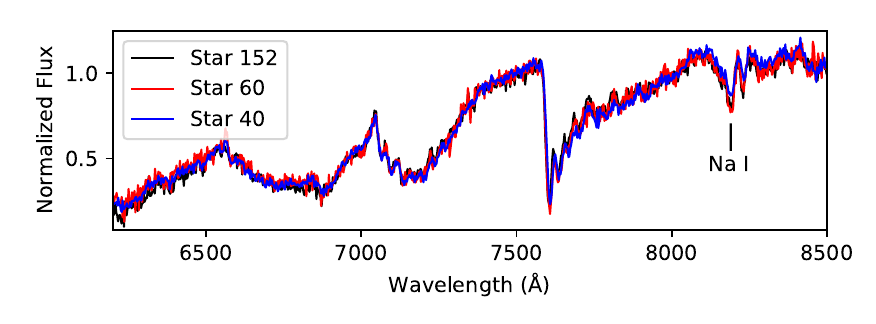}
  \caption{The spectra of Star \#40 (GWAC201113A), \#60 (GWAC191030A) and \#152 (GWAC180212A). The spectra are normalized by dividing the fluxes at 7500\,\AA.}\label{fig:specmp}
\end{figure}

Both Star \#40 (GWAC201218A) and \#60 (GWAC191030A) have a spectral type of M3.5, but their $G_{\rm BP}- G_{\rm RP}$ are in the range of M2  as shown in Figure \ref{fig:cmd}. Their spectra were obtained with the 2.16m Telescope and are plotted in Figure \ref{fig:specmp} with the comparing spectrum of the normal M3.5V Star \#152 (GWAC180212A), which was also obtained with the 2.16m Telescope. From their spectra we can see that spectra of these 3 stars are similar, but the Na I doublet line around 8190 \AA \, of  Star \#40 (GWAC201218A) is significantly weaker than those of Star \#60 (GWAC191030A) and \#152 (GWAC180212A), which means that Star \#60 (GWAC191030A) is a dwarf, while Star \#40 (GWAC201218A) is a giant or young star. We checked Star \#40 (GWAC201218A) and noticed that its $ruwe = 6.958$ in Gaia DR3\footnote{\url{https://gea.esac.esa.int/archive/documentation/GDR2/Gaia_archive/chap_datamodel/sec_dm_main_tables/ssec_dm_ruwe.html}}, which implies its data may be unreliable.  As for Star \#60 (GWAC191030A), it is still unclear why its $G_{\rm BP}- G_{\rm RP}$ is significantly bluer than other M3.5V stars in Figure \ref{fig:cmd}.
\par

Star \#59 (GWAC210109A) is an M5.5 young star in the Taurus star-forming region \citep{2019AJ....158...54E}.  Star \#155 (GWAC211016A) is contaminated by a nearby star with a distance of $<1.''2$, so its Gaia data may be unreliable. 
\par
\begin{figure}
  \centering
  \includegraphics[scale=0.5]{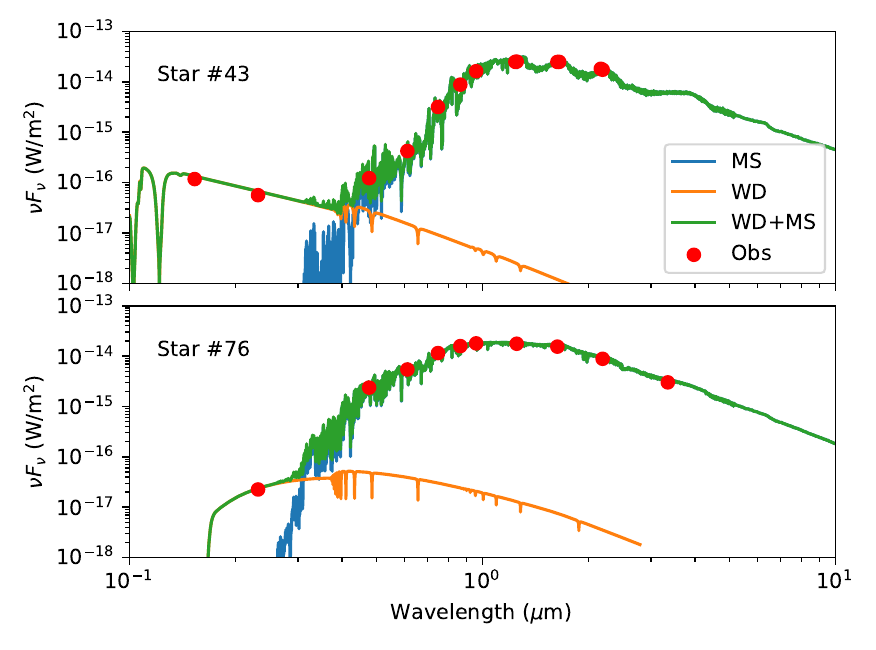}
  \caption{The SED fittings of Star \#43 (GWAC190121B) and \#76 (GWAC201113A). In each panel, the blue spectrum is the template for a main sequence star (MS), the yellow spectrum is a template for a white dwarf star (WD), the green spectrum is the WD+MS spectrum, and the red circles are the photometries of GALEX FUV, NUV, PS1 g, r, i, z, y, 2MASS J, H, K and WISE W1, W2 bands. } \label{fig:sed}
\end{figure}

Star \#143 (GWAC210518A) was reported to be a white dwarf + main sequence  (WDMS) binary \citep{ 2012MNRAS.419..806R}. To find more WDMS stars, for Star \#34 (GWAC190901A), \#43 (GWAC190121B) and \#75 (GWAC190324A), \#76 (GWAC201113A), \#98 (GWAC180117A), \#102 (GWAC190206A), \#106 (GWAC201216A), \#117 (GWAC190116B), \#133 (GWAC200514A) and \#135 (GWAC200319A) with abnormally higher FUV and NUV, we fitted their SEDs of {\it GALEX} FUV, NUV, PS1 g, r, i, z, y, 2MASS J, H, K and {\it WISE} \citep{2010AJ....140.1868W} W1, W2 bands by a white dwarf and a main sequence star templates \citep{2023AJ....165..119Y}. Finally, only Star \#43 (GWAC190121B) and \#76 (GWAC201113A) seem to be well fitted as shown in Figure \ref{fig:sed}. Certainly, UV excesses of these stars may be from their active chromospheres. 
\par
For stars that have no available spectra, their spectral types were assigned by their $G_{\rm BP} - G_{\rm RP}$ in Figure \ref{fig:cmd}.
From Figure \ref{fig:cmd}, we can see that spectral types of GWAC flare stars range from M2 to L1, and most stars are M4, which implies that these stars around the convective boundary tend to produce big flares.
\par 
The PARSEC isochrones \citep{2014MNRAS.444.2525C} of 0.01 Gyr, 0.1 Gyr and 1 Gyr with [Fe/H] $=$ 0 are overplotted in Figure \ref{fig:cmd}, from which we can see that most stars should be younger than 1 Gyr.

\subsection{Flare Frequency Distribution} \label{subsec:ffd}

\begin{figure}
  \centering
  \includegraphics[scale=0.6]{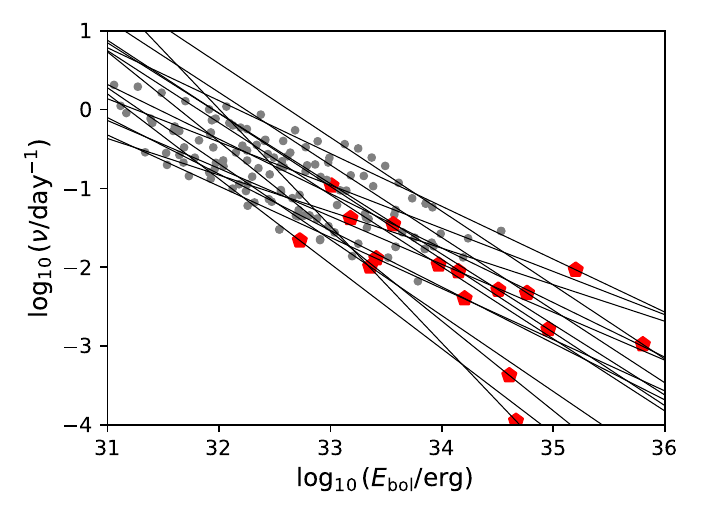}
  \caption{The cumulative flare frequency distributions (FFDs) for 16 stars with more than 20 flares detected in TESS or K2 light curves. The cumulative flare frequencies are denoted by gray circles, and black lines are the fitted functions; The bolometric energies of GWAC flares are shown by red pentagons.  } 
  \label{fig:ffd}
\end{figure}

There are 16 GWAC flare stars with more than 20 flares in their TESS or K2 light curves, and their cumulative flare frequency distributions (FFDs) are shown in Figure \ref{fig:ffd}, where the cumulative flare frequencies of each star are shown in gray circles with the bolometric flare energies from a blackbody of $T=9000$ K. 
Each FFD is fitted by the linear function 
\begin{equation}
\label{equ:ffd}
\log_{10} ({\rm \nu}) = \alpha \log_{10}(E_{\rm bol}) + \beta
\end{equation}
 and shown by a black line. Here, $\nu$ is the cumulative flare frequency in day$^{-1}$, and $\alpha$ and $\beta$ are the parameters to be fitted. These 16 stars with their  $\alpha$ and $\beta$ and the predicted frequencies of GWAC flares by Equation \ref{equ:ffd} are shown in Table \ref{tab:ffd}.
The GWAC flare energies on the fitted FFD lines are shown by red pentagons in Figure \ref{fig:ffd}.  We found that most (13/16) GWAC flares can happen more than once every year, and only three GWAC flares happen once every several or even 20 yr.

\par
  
\begin{deluxetable*}{lllllllll}
		\tablecaption{Stars with More than 20 Flares \label{tab:ffd}}
		\tablewidth{700pt}
		\tabletypesize{\scriptsize}
		\tablehead{ \colhead{Star No.} & \colhead{TIC/K2} & \colhead{SpT} & \colhead{$\alpha$} & \colhead{$\alpha$\_err} & \colhead{$\beta$} & \colhead{$\beta$\_err}& \colhead{$\log_{10}\nu_{\rm GWAC}$} & \colhead{$\log_{10}\nu_{\rm GWAC}$\_err} \\
		 \colhead{} & \colhead{} & \colhead{} & \colhead{$\left[ {\rm day}^{-1} \right]$} &\colhead{$\left[ {\rm day}^{-1} \right]$} & \colhead{$\left[ {\rm day}^{-1} \right]$} & \colhead{$\left[ {\rm day}^{-1} \right]$}&\colhead{$\left[ {\rm day}^{-1} \right]$}&\colhead{$\left[ {\rm day}^{-1} \right]$} 
	}
\startdata
11&EPIC 220404032&M4V&-0.6705&0.0294&21.57&0.98&-2.03&0.06\\
30&TIC 22944327&M4.5V&-0.4635&0.0246&14.00&0.81&-1.38&0.05\\
33&TIC 328254412&M5.5V&-0.5478&0.0262&17.12&0.86&-0.96&0.05\\
44&EPIC 210878635&M4V&-0.8935&0.0530&28.55&1.74&-2.29&0.07\\
49&EPIC 210417498&M4V&-1.1381&0.1041&36.01&3.38&-3.37&0.07\\
52&TIC 283866910&M4.5V&-0.9140&0.0745&29.08&2.39&-1.97&0.06\\
56&TIC 245936201&M3.5V&-0.9368&0.0993&30.56&3.32&-2.98&0.08\\
77&TIC 461654150&M8V&-0.6492&0.0423&19.80&1.36&-1.89&0.05\\
86&EPIC 212002525&M4.5V&-0.9269&0.0563&29.61&1.86&-2.79&0.07\\
87&TIC 175241416&M7V&-0.9632&0.0563&30.13&1.80&-1.99&0.08\\
93&TIC 471012520&M6V&-1.0777&0.0588&33.61&1.87&-1.66&0.09\\
96&TIC 251079483&M4.5V&-0.6915&0.0545&21.75&1.78&-1.46&0.09\\
113&TIC 289534997&M6V&-0.7162&0.0219&22.10&0.72&-2.40&0.04\\
117&EPIC 248853090&M4.5V&-0.9209&0.0426&29.69&1.42&-2.33&0.06\\
119&TIC 156151200&M5V&-0.6080&0.0332&18.71&1.08&-2.05&0.06\\
120&EPIC 201664337&M4V&-1.4847&0.1240&47.52&4.04&-3.95&0.13\\
\enddata
\tablecomments{
	$\alpha$ and $\beta$ are the parameters in Equation \ref{equ:ffd}, and $\alpha$\_err and $\beta$\_err are respectively their errors.  $\log_{10}\nu_{\rm GWAC}$ is the logarithm of the predicted frequency of the GWAC flare energy with the error of $\log_{10}\nu_{\rm GWAC}$\_err.
		}
\end{deluxetable*}

\subsection{Kinematics}\label{subsec:vel}
\begin{figure}
  \centering
   \includegraphics[scale=0.75]{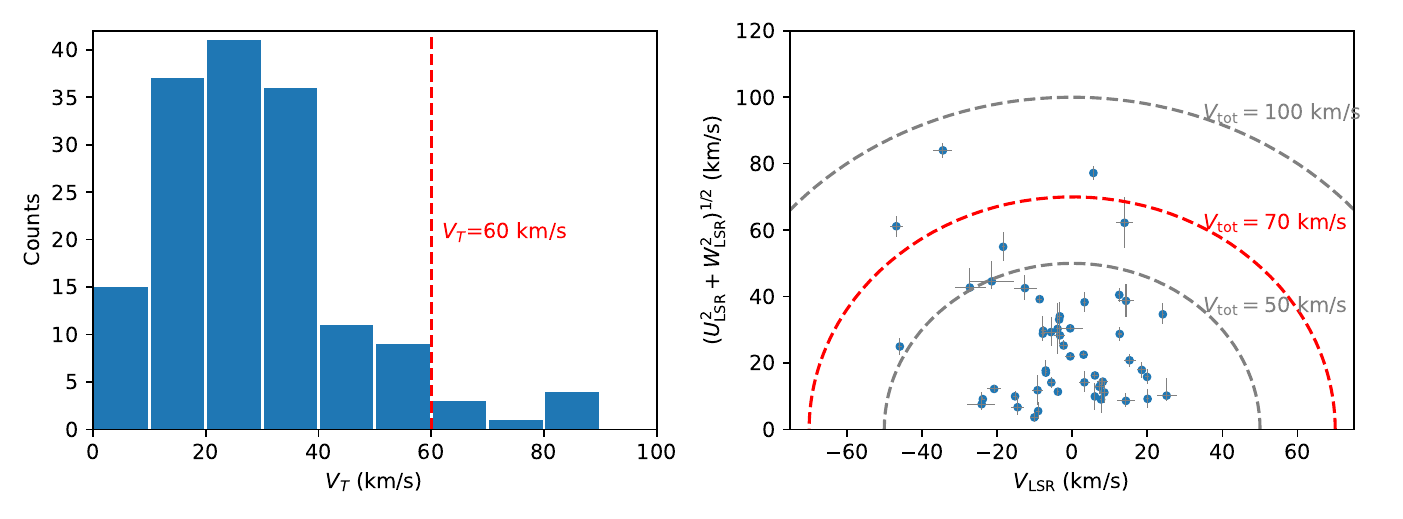}
  \caption{Left panel: the distributions of tangential velocities of 157 GWAC flare stars; Right panel: Toomre diagram for 52 GWAC flare stars with available radial velocities. The 1$\sigma$ errors are shown in gray lines. } 
   \label{fig:vel}
\end{figure}

To explore kinematic properties of these flare stars, we calculated their tangential velocities ($V_{\rm T}$) for stars with available proper motions and parallaxes in Gaia DR3, and for stars that are not proven to be binaries in this work and have reliable radial velocities in SDSS and LAMOST data, their velocities in the local standard of rest (LSR) ($U_{\rm LSR}, V_{\rm LSR}, W_{\rm LSR}$) were also calculated by the python package \emph{astropy.coordinates.SkyCoord} with the velocity of the Sun relative to the LSR of  $(U_{\rm LSR}, V_{\rm LSR}, W_{\rm LSR})_{\odot} = (11.1, 12.24, 7.25)$ km\,s$^{-1}$ \citep{2010MNRAS.403.1829S}. We notice that there are some stars have several available radial velocities, so the medium velocity was used during calculation.  Here, $U_{\rm LSR}$ is directed to the Galactic center, $V_{\rm LSR}$ is in the Galactic rotational direction, and $W_{\rm LSR}$ points to the Galactic North Pole. The $V_{\rm T}$ was calculated by the proper motion $\mu$ and parallax $\varpi$ in Gaia DR3: $V_{\rm T} = 4.74 \mu/\varpi$, and the total velocity was calculated by $V_{tot} = (U_{\rm LSR}^2 + V_{\rm LSR}^2 + W_{\rm LSR}^2)^{1/2}$. As a result, there are 157 stars have $V_{\rm T}$. After removing spectroscopic binaries, there are 52 stars have velocities relative to the LSR. The results are shown in Figure \ref{fig:vel}.
\par
We used the certiria in \citet{2018A&A...616A..10G} to select thin disk ($V_{\rm T}<40$ km\,s$^{-1}$  or $V_{\rm tot}<50$ km\,s$^{-1}$), thick disk ($60<V_T<150$ km\,s$^{-1}$ or $70<V_{\rm tot}<180$ km\,s$^{-1}$) and halo stars ( $V_{\rm T}>200$ km\,s$^{-1}$ or  $V_{\rm tot}>200$ km\,s$^{-1}$ ).  From Figure \ref{fig:vel} we can see that there is no halo star, and only 8/157 ($V_{\rm T}>60$ km\,s$^{-2}$) or 3/52 ($V_{\rm tot}>70$ km\,s$^{-2}$) are thick disk stars, but speeds of these stars are far from the upper limit of thick disk stars. Therefore, GWAC flare stars are not old and most of them should be young. 

\section{Rotation-Age-Activity Relation} \label{sec:rot}
The periods of 105/162 stars were obtained from their TESS or K2 light curves by the Lomb-Scargle method. We inspected all of the light curves by eye and found there are no EA and EB eclipses, so the periods should be rotational periods. We also inspected all folded light curves by eye and found that most of them are sinusoidal, so if there are some EW binaries in GWAC flare stars, their real periods would be halved.
\par
\begin{figure}
  \centering
  \includegraphics[scale=0.5]{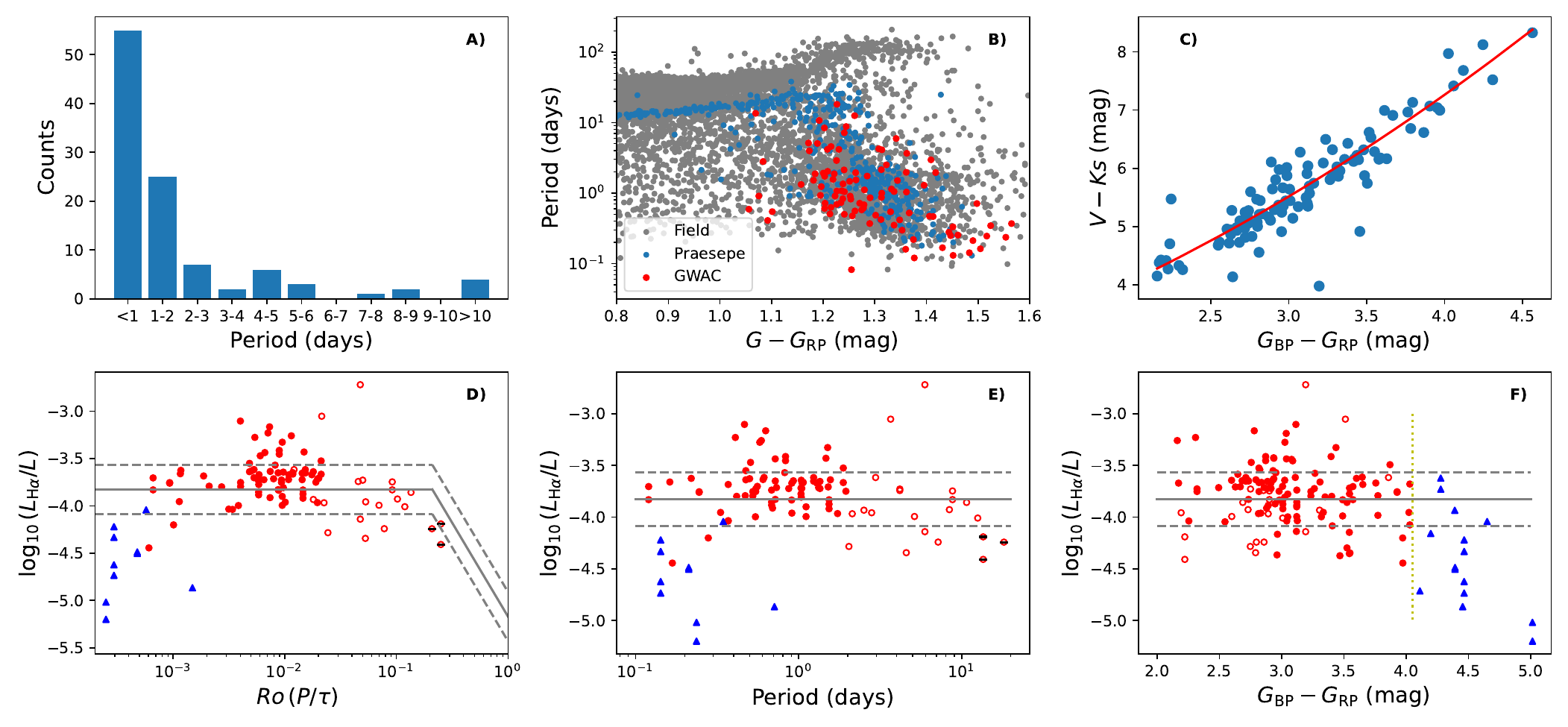}
  \caption{The rotation-age-activity relation. Panel A: The distribution of periods of 105 GWAC flare stars.  Panel B: The color-period diagram. The gray and blue circles are, respectively, field stars and Praesepe stars from \citet{2021ApJ...916...77P}, and the red  circles are GWAC flare stars. Panel C: $V - Ks$ vs. $G_{\rm BP} - G_{\rm RP}$. The fitted relationship (Equation \ref{equ:gv}) is shown in the red curve. Panel D:  $\log_{10}(L_{\rm H\alpha}/L_{\rm bol})$ vs. Rossby number ($Ro$) for stars with available spectra and $Ro$. Panel E:  $\log_{10}(L_{\rm H\alpha}/L_{\rm bol})$ vs. period for stars with available spectra and periods. Panel F:  The $\log_{10}(L_{\rm H\alpha}/L_{\rm bol})$ vs. $G_{\rm BP} - G_{\rm RP}$ diagram for 117 GWAC stars that have available spectra. The dotted yellow vertical line is $G_{\rm BP} - G_{\rm RP} = 4.05$, which is roughly the demarcation for M6 and M7 stars. In Panel D, E, and F, the blue triangles are M7-L1 stars and the red circles are stars M2-M6 stars. For red circles, the filled ones are stars with $ P < 2$ days and the empty ones are stars with $P> 2$ days. For stars that have more than one available $\log_{10}(L_{\rm H\alpha}/L_{\rm bol})$, all  $\log_{10}(L_{\rm H\alpha}/L_{\rm bol})$ are shown. The central gray line is the saturation line ($L_{\rm H\alpha}/L = 1.49\times 10^{-4}$) and gray dotted lines are the 1$\sigma$ ($\sigma=0.26$) positions given in \citet{2017ApJ...834...85N}. The black horizontal short lines are Star \#51 and \#56, which are around the saturation threshold. }
   \label{fig:pd-age}
\end{figure}

 The periods are shown in Panel A of Figure \ref{fig:pd-age}. Most stars (80/105) have periods shorter than 2 days, and only 4 stars have periods longer than 10 days. 
In Panel B, GWAC flare stars are shown in red circles in the color-period diagram. The gray and blue circles are, respectively, field and Praesepe stars from \citet{2021ApJ...916...77P}. The gray circles show that the most field stars converge to the upper belt with periods greater than several 10 days. Praesepe has an age of $\sim$670 Myr, and its upper periods are about 10-20 days. There is a fast reservoir of $1.2< G-G_{\rm RP} < 1.4$ and $0.2 < P < 2$ days \citep{2021ApJ...916...77P}, where GWAC flare stars with available periods are also clustered. From this panel we can see that GWAC flare stars are still far from the upper period belt where the field stars cluster, and rotate even faster than Praesepe stars,  so all these 105/162 GWAC stars should be younger than 670 Myr. 
\par
In the works of \citet{2018MNRAS.479.2351W} and \citet{2011ApJ...743...48W}, the convective turnover time $\tau$ is a function of $V-Ks$, but the GWAC flare stars are very red and faint, and then some of them have no available $V$ data. As a result, we firstly obtained $V - Ks$ of some GWAC flare stars from \citet{2022yCat.4039....0P}, then fitted the relation of $V-Ks$ vs. $G_{\rm BP} -G_{\rm RP}$, and finally obtained their predicted $V-Ks$ from $G_{\rm BP} - G_{\rm RP}$:
\begin{equation}
 V-Ks = 0.15226\times (G_{\rm BP} - G_{\rm RP})^2 +0.67908\times (G_{\rm BP} - G_{\rm RP}) + 2.1097 
 \label{equ:gv}
 \end{equation}
which is shown in Panel C. Then we used the function given by \citet{2018MNRAS.479.2351W} to calculate the convective turnover time $\tau$:
$$\log_{10}(\tau) = 0.64 + 0.25 \times (V - Ks )$$.
\par

There are 108 spectra with available $\log_{10}(L_{\rm H\alpha}/L_{\rm bol})$ for 82 GWAC flare stars, which also have available periods. 
Their $\log_{10}(L_{\rm H\alpha}/L_{\rm bol})$ vs. Rossby number ($Ro=P/\tau$) diagram is shown in Panel D. For stars that have more than one available $\log_{10}(L_{\rm H\alpha}/L_{\rm bol})$, all $\log_{10}(L_{\rm H\alpha}/L_{\rm bol})$ are shown. 
We also used the equation presented by \citet{2014ApJ...794..144R} to calculate the saturation period $P_{\rm sat}$ in days:
\begin{equation}\label{equ:sat}
P_{\rm sat} =(1.1\times 10^{34}/L_{\rm bol})^{1/2}
\end{equation}
We found that only Star \#51 and \#56 are around the saturation $Ro$ and also $P_{\rm sat}$ which are shown in Panel D and E of Figure \ref{fig:pd-age} by black horizontal lines, and all other stars are in the saturation region.
There are 117 GWAC flare stars have available $L_{\rm H\alpha}$ in total, and their  $\log_{10}(L_{\rm H\alpha}/L_{\rm bol})$ vs. $G_{\rm BP} -G_{\rm RP}$ diagram is shown in Panel F. In Panel D, E, and F, the blue triangles are stars of M7-L1 and the red circles are stars of M2-M6. For red circles, the filled ones are stars with $ P < 2$ days and the empty ones are stars with $P> 2$ days. The central gray line is the saturation line ($L_{\rm H\alpha}/L = 1.49\times 10^{-4}$) and gray dotted lines are the 1$\sigma$ ($\sigma=0.26$) positions given in \citet{2017ApJ...834...85N}. 
From Panel D, E and F, we can see that: 
\begin{itemize}
\item Stars in Panel D and E are all in or very near to the saturation region, so these stars should be very active.
\item Though in the saturation region, $\log_{10}(L_{\rm H\alpha}/L_{\rm bol})$ are significantly lower for M7 - L1 stars (blue triangles) than for M2 - M6 stars (red circles). 
\item For M2 - M6 stars (red circles), it seems that $L_{\rm H\alpha}/L_{\rm bol}$ slightly decreases with increasing $Ro$ (as seen in Panel D) and also increasing period (as seen in Panel E), but is independent of $G_{\rm BP} - G_{\rm RP}$ (as seen in Panel F). Especially, the $L_{\rm H\alpha}/L_{\rm bol}$ are lower for stars with $P>2$ days (red empty circles) than for those with $P<2$ days (red filled circles). This declination of  $\log_{10}(L_{\rm H\alpha}/L_{\rm bol})$ with increasing period may imply the increasing stellar age \citep{2021AJ....161..277K}.
\item As shown in Panel F, $\log_{10}(L_{\rm H\alpha}/L_{\rm bol})$ seems to decrease with increasing $G_{\rm BP} - G_{\rm RP}$ (i.e. spectral type), which was also presented by \citet{2010ApJ...709..332B} and \citet{2015AJ....149..158S}.
\end{itemize}
In summary, for M2 - M6 stars, their  $\log_{10}(L_{\rm H\alpha}/L_{\rm bol})$ decreases with increasing period and is independent of spectral type, but for M7 - L1 stars, though they are rapid rotators, their  $\log_{10}(L_{\rm H\alpha}/L_{\rm bol})$ are significantly lower than those of M2 - M6 stars.
\par
The functions to calculate the convective overturn time given by \citet{2018MNRAS.479.2351W,2011ApJ...743...48W} and \citet{2014ApJ...794..144R} are frequently used in literature, but these equations were obtained from the stars with masses larger than 0.1 M$_{\odot}$ or spectral types no later than M6 \citep{2018MNRAS.479.2351W,2011ApJ...743...48W,2022ApJ...940..145J}. Therefore, we do not know if Ro and saturation periods derived from these functions can apply to stars later than M6. However, in any case, it is true that the $\log_{10}(L_{\rm H\alpha}/L_{\rm bol})$ is lower for M7 - L0 stars than for M2 - M6 stars as shown in Panel D, E, F of Figure \ref{fig:pd-age}, because it is also seen in literature \citep[e.g.][]{2010ApJ...709..332B,2015AJ....149..158S,2021AJ....161..277K}.

\section{Maximum Flare Energy} \label{sec:maxerg}
\begin{figure}
  \centering
  \includegraphics[scale=0.8]{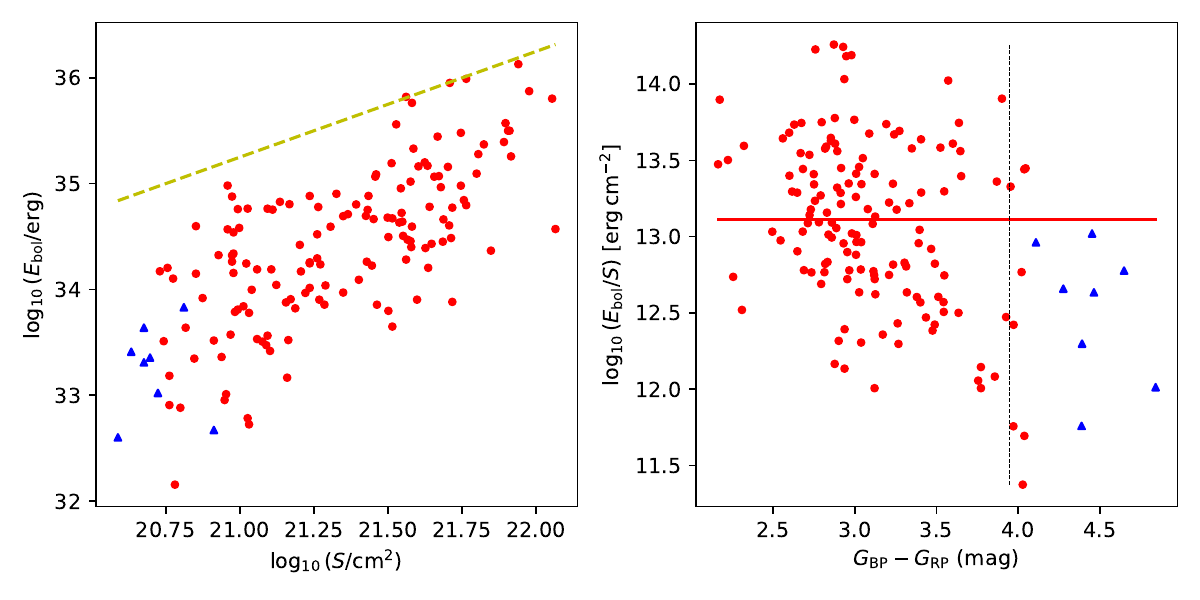}
  \caption{ The $E_{\rm bol}$-$S$-$T_{\rm eff}$ relation. The red circles are M2-M6 stars, and the blue triangles are M7-L1 stars. The yellow dashed line in the right panel is the lower limit of the maximum flare energy, which is $\log_{10}E_{\rm bol} = \log_{10}S + 14.25$. In the right panel, the vertical dotted line is $G_{\rm BP} - G_{\rm RP} = 3.95$ and roughly the demarcation between M2-M6 and M7-L1 stars, and the red horizontal line is the median value of $\log_{10}(E_{\rm bol}/S)$ of M2-M6 stars. } 
  \label{fig:e/s}
\end{figure}

Some works \citep[e.g.][]{2019ApJ...876...58N,2013PASJ...65...49S} suggested that for a solar-like star, the upper limit of flare energy is usually determined by the spot size, and thus the stellar hemisphere. 
Inspired by this, we inspected the relation of the flare bolometric energy  $E_{\rm bol}$ vs. the area of stellar hemisphere $S = 2\pi R^2$ ($R$ is the stellar radius from TIC 8.2 ), as shown in the left panel of Figure \ref{fig:e/s}. From this panel we can see that $E_{\rm bol}$ increases with increasing $S$. The relation of flare energy divided by stellar hemisphere area $\log_{10}(E_{\rm bol}/S)$ vs. stellar surface effective temperature $T_{\rm eff}$ ($T_{\rm eff}$ from TIC 8.2), is shown in the right panel of Figure \ref{fig:e/s}, where the vertical dashed line is  $G_{\rm BP} - G_{\rm RP} = 3.95$ and roughly the demarcation between M2 - M6 and M7-L1 stars, and the red horizontal line is the median value of $\log_{10}(E_{\rm bol}/S)$ of M2-M6 stars. From this panel we can see that for M2-M6 stars, the distribution of $\log_{10}(E_{\rm bol}/S)$ is roughly same and independent of $T_{\rm eff}$, and significantly higher than those of M7-L1 stars. We found that the maximum $\log_{10}(E_{\rm bol}/S)$ is about 14.25 for GWAC flares in the right panel, which means that for M2-M6 stars, the lower limit of the maximum flare energy is
\begin{equation}
\log_{10}E_{\rm bol} = \log_{10}S + 14.25
\label{equ:me}
\end{equation}
which is also shown by the yellow dashed line in the left panel of Figure \ref{fig:e/s}.
\par

\section{Discussion} \label{sec:dis} 
\subsection{Stellar Age} \label{subsec:age}
There are 105/162 stars having periods as shown in Panel A of Figure \ref{fig:pd-age}. \citet{2022ApJ...935..104M} suggests that for stars with $P_{\rm rot} < 10$ days, their ages are from about 600 Myr for 0.2 - 0.3 M$_{\odot}$ to about 2 - 3 Gyr for 0.1-0.2 M$_{\odot}$. \citet{2022ApJ...936..109P} also suggested that for stars with 0.2 - 0.3 M$_{\odot}$ and $P_{\rm rot}<2$ days, their ages are less than 600 Myr, and $2<P_{\rm rot}<10$ days means their ages between 1-3 Gyr. Therefore, these GWAC stars should be younger than 600 Myr - 3 Gyr. By comparing with Praesepe stars in Panel B of Figure \ref{fig:pd-age}, we refined their ages as younger than 670 Myr.  
\par
There are 38 stars having available $L_{\rm H\alpha}/L_{\rm bol}$ but without available periods.
 \citet{2021AJ....161..277K} proposed (see their Figure 8) that the activity of M-type stars decreases with age. In their Figure 6, for M2 - M6 stars younger than 1 Gyr, their $L_{\rm H\alpha}/L_{\rm bol}$ would significantly greater than $10^{-4}$, and then decreases abruptly to around $10^{-4.25} \approx 5.6\times 10^{-5}$ at 1 - 3 Gyr, while for stars of M7 - M9,  $L_{\rm H\alpha}/L_{\rm bol}$ decreases abruptly to significantly $<10^{-4.5} \approx 3.2 \times 10^{-5}$ during $\sim$ 30 Myr through 3 Gyr. As a result, for stars in Panel F of Figure \ref{fig:pd-age}, M2-M6 and M7-L1 stars should be younger than 1 and 3 Gyr, respectively. 
 \par
 Among the 157 stars with $V_{\rm T} > 60$ km\,s$^{-1}$ in the left panel of Figure \ref{fig:vel}, Star \#40, \#132, and \#145 have neither an available period nor $L_{\rm H\alpha}/L_{\rm bol}$. For the remaining five stars without available $V_{\rm T}$, only Star \#19 has neither available period nor $L_{\rm H\alpha}/L_{\rm bol}$. Therefore, there are only 4/162 GWAC flare stars that cannot be determined if they belong to the thin disk. Therefore, except for these four stars, all other 158/162 stars should be young.
 \par
In summary, if we think these GWAC flare stars are from the same category, then their periods show they are younger than 670 Myr, $L_{\rm H\alpha}/L_{\rm bol}$ show they are younger than 3 Gyr, and $V_{\rm T}$ shows they are the thin disk stars.
 
\subsection{The Lower Activity Level of stars later than M6} \label{subsec:twoact}

For M and L stars, $L_{\rm H\alpha}/L_{\rm bol}$ decreases with increasing spectral type \citep{2003ApJ...583..451M,2010ApJ...709..332B,2015AJ....149..158S}, which is also shown in Panel F of Figure \ref{fig:pd-age}, and the $L_{\rm H\alpha}/L_{\rm bol}$ level is significantly lower for stars later than M6 than for earlier type stars.
\citet{2021AJ....161..277K} suggested that for M7-M9 stars, they have  the same rotation-activity relation as earlier type stars only in their first $\sim$30 Myr, and after that their activity decreases abruptly and keeps at the low level for more than 3-6 Gyr. 
\par
Besides $L_{\rm H\alpha}/L_{\rm bol}$, \citet{2010ApJ...709..332B} also found that for the stellar activity there is a breakout around M7 in the quiescent X-ray and radio:
for stars ranging from F to M6, they have the similar $L_{\rm X}/L_{\rm bol}$ pattern and $L_{\rm X}/L_{\rm bol}$ is higher than that of later type stars; $L_{\rm radio}/L_{\rm bol}$ and $L_{\rm radio}/L_{\rm X}$ are lower for M1-M6 stars than for later type stars.
\par
In the GWAC sample, for M7-L1 stars,  the distributions of their $\bigtriangleup G$ and $ED$ are similar to those of M2-M6 stars as shown in Figure \ref{fig:amp}, but their $\log_{10}(E_{\rm bol}/S)$ are lower than the median $\log_{10}(E_{\rm bol}/S)$ of  M2 - M6 stars as shown in the right panel of Figure \ref{fig:e/s}. Therefore, if the stellar radii given in TIC 8.2 are reliable, then the $\log_{10}(E_{\rm bol}/S)$ are lower for M7 - L1 stars than for M2-M6 stars. However, the number of M7-L1 stars is too small to draw a firm conclusion. 
 \par
The lower activity level of later type stars has been noticed for a long time. \citet{2002ApJ...571..469M} suggested that as the surface temperature decreases, the resistivity of the stellar atmosphere increases, then the charged particles in the magnetic field are carried away due to collisions with neutral gases and thus the magnetic free energy used to produce the flare decreases with the spectral type. However, this theory still needs to be examined by further observations.
\section{Conclusion} \label{sec:con}
In this work, we presented 163 big flares in GWAC triggers from 162 individual stars with the spectral type from M2 to L1. The flare amplitude $\bigtriangleup G$ ranges from 0.83 to $\sim$10 mag, and the flare energy ranges from $10^{32.2}$ to $10^{36.4}$ erg. From TESS or K2 light curves, we found 1478 flares from 117 stars and calculated periods for 105 stars. Besides, we obtained 154 low-resolution spectra for 120 individual stars with the 2.16 m Telescope in the Xinglong Station, LAMOST and SDSS. Among them, there are 108 available $L_{\rm H\alpha}/L_{\rm bol}$ for 82 individual stars with available periods. We also obtained tangent velocities $V_{\rm T}$ for 157 stars , and velocities relative to LSR for 52 stars.
%
From these data we found that:
\begin{itemize}
\item The energy of GWAC flares decreases with increasing $G_{\rm BP} - G_{\rm RP}$ (decreasing stellar surface effective temperature $T_{\rm eff}$) (see Figure \ref{fig:erg}), but both distributions of flare amplitude $\bigtriangleup G$ and flare equivalent duration  $\log_{10}(ED)$ seem to be independent of $T_{\rm eff}$ (see Figure \ref{fig:amp}).
\item If these GWAC flare stars are from the same category,  then their periods show they are younger than 670 Myr, $L_{\rm H\alpha}/L_{\rm bol}$ show they are younger than 3 Gyr, and $V_{\rm T}$ show they are the thin disk stars (see Figure \ref{fig:vel} and \ref{fig:pd-age}). Therefore, they are young stars.
\item There are 16 stars with flare numbers greater than 20. From their FFDs we found that most (13/16) GWAC flares can happen more than once every year, and only three GWAC flares happen once every several or even 20 years (see Figure \ref{fig:ffd}).
\item Stars with spectra and periods are in or very near to  the saturation region, so these stars should be very active.
\item $\log_{10}(L_{\rm H\alpha}/L_{\rm bol})$ is higher for M2-M6 stars than for M7-L1 stars.  $\log_{10}(L_{\rm H\alpha}/L_{\rm bol})$ of M2-M6 stars is around the saturation level, and decreases with increasing period, while for M7-L1 stars, though they are all fast rotators, their $L_{\rm H\alpha}/L_{\rm bol}$ are significantly lower (see Figure \ref{fig:pd-age}).
\item  The flare energy divided by the stellar hemispherical area $\log_{10}(E_{\rm bol}/S)$ seems to be higher for M2 - M6 stars than for M7-L1 stars   (see Figure \ref{fig:e/s}). However, the number of M7-L1 stars is too small to draw a firm conclusion.

\item The maximum flare energy of M2 - M6 stars should be larger than $\log_{10}E_{\rm bol} = \log_{10}S + 14.25$ (see Figure \ref{fig:e/s}).
\end{itemize}

\par
The different activity levels between  M7-L1 and M2-M6 stars had been reported by \citet{2010ApJ...709..332B} in X-ray, radio and H$\alpha$, but we still do not know if the different activity levels exist in  $\log_{10}(E_{\rm bol}/S)$, which should be explored by the sample including a lot of stars later than M6 that release big white-light flare energies. The study would be very interesting because the origin of the activity pattern of cool stars is still unclear.
%

\acknowledgments
The authors thank the anonymous referee very much for the valuable report that inspired us to improve this work. 
This work is supported by the National Natural Science Foundation of China (NSFC) with grant No. 12073038. Liang Wang acknowledges the National Natural Science Foundation of China under Grant No. U2031144. Hai-Long Yuan acknowledges the support from the Youth Innovation Promotion Association of the CAS (Id. 2020060). We acknowledge the science research grants from the China Manned Space Project.

\par
Guoshoujing Telescope (the Large Sky Area Multi-Object Fiber Spectroscopic Telescope LAMOST) is a National Major Scientific Project built by the Chinese Academy of Sciences, Funding for the project has been provided by the National Development and Reform Commission. LAMOST is operated and managed by the National Astronomical Observatories, the Chinese Academy of Sciences.
\par
We acknowledge the support of the staff of the Xinglong 2.16m telescope. This work was partially supported by the Open Project Program of the Key Laboratory of Optical Astronomy, National Astronomical Observatories, Chinese Academy of Sciences. 
\par
This research has made use of "Aladin sky atlas" developed at CDS, Strasbourg Observatory, France.
\par

This work has made use of data from the European Space Agency (ESA) mission
{\it Gaia} (\url{https://www.cosmos.esa.int/gaia}), processed by the {\it Gaia}
Data Processing and Analysis Consortium (DPAC,
\url{https://www.cosmos.esa.int/web/gaia/dpac/consortium}). Funding for the DPAC
has been provided by national institutions, in particular the institutions
participating in the {\it Gaia} Multilateral Agreement.
\par
Some/all of the data presented in this paper were obtained from the Mikulski Archive for Space Telescopes (MAST) at the Space Telescope Science Institute. The specific observations analyzed can be accessed via \dataset[https://doi.org/10.17909/55e7-5x63]{https://doi.org/10.17909/55e7-5x63} \citep{https://doi.org/10.17909/55e7-5x63}, \dataset[https://doi.org/10.17909/T9H59D]{https://doi.org/10.17909/T9H59D} \citep{https://doi.org/10.17909/T9H59D}, \dataset[https://doi.org/10.17909/fwdt-2x66]{https://doi.org/10.17909/fwdt-2x66} \citep{https://doi.org/10.17909/fwdt-2x66}, \dataset[https://doi.org/10.17909/T93W28]{https://doi.org/10.17909/T93W28} \citep{https://doi.org/10.17909/T93W28}. STScI is operated by the Association of Universities for Research in Astronomy, Inc., under NASA contract NAS5–26555. Support to MAST for these data is provided by the NASA Office of Space Science via grant NAG5–7584 and by other grants and contracts.
\par
This publication makes use of data products from the Wide-field Infrared Survey Explorer, which is a joint project of the University of California, Los Angeles, and the Jet Propulsion Laboratory/California Institute of Technology, funded by the National Aeronautics and Space Administration.
\par
This publication makes use of data products from the Two Micron All Sky Survey, which is a joint project of the University of Massachusetts and the Infrared Processing and Analysis Center/California Institute of Technology, funded by the National Aeronautics and Space Administration and the National Science Foundation.

%

\vspace{5mm}


\software{astropy \citep{2013A&A...558A..33A},  lightkurve \citep{2018ascl.soft12013L}
          }




\bibliography{superflare}{}

\begin{thebibliography}{}
\expandafter\ifx\csname natexlab\endcsname\relax\def\natexlab#1{#1}\fi

\bibitem[{{Althukair} \& {Tsiklauri}(2023)}]{2023RAA....23h5017A}
{Althukair}, A.~K., \& {Tsiklauri}, D. 2023, Research in Astronomy and
  Astrophysics, 23, 085017

\bibitem[{{Astropy Collaboration} {et~al.}(2013){Astropy Collaboration},
  {Robitaille}, {Tollerud}, {Greenfield}, {Droettboom}, {Bray}, {Aldcroft},
  {Davis}, {Ginsburg}, {Price-Whelan}, {Kerzendorf}, {Conley}, {Crighton},
  {Barbary}, {Muna}, {Ferguson}, {Grollier}, {Parikh}, {Nair}, {Unther},
  {Deil}, {Woillez}, {Conseil}, {Kramer}, {Turner}, {Singer}, {Fox}, {Weaver},
  {Zabalza}, {Edwards}, {Azalee Bostroem}, {Burke}, {Casey}, {Crawford},
  {Dencheva}, {Ely}, {Jenness}, {Labrie}, {Lim}, {Pierfederici}, {Pontzen},
  {Ptak}, {Refsdal}, {Servillat}, \& {Streicher}}]{2013A&A...558A..33A}
{Astropy Collaboration}, {Robitaille}, T.~P., {Tollerud}, E.~J., {et~al.} 2013,
  \aap, 558, A33

\bibitem[{{Astropy Collaboration} {et~al.}(2018){Astropy Collaboration},
  {Price-Whelan}, {Sip{\H{o}}cz}, {G{\"u}nther}, {Lim}, {Crawford}, {Conseil},
  {Shupe}, {Craig}, {Dencheva}, {Ginsburg}, {VanderPlas}, {Bradley},
  {P{\'e}rez-Su{\'a}rez}, {de Val-Borro}, {Aldcroft}, {Cruz}, {Robitaille},
  {Tollerud}, {Ardelean}, {Babej}, {Bach}, {Bachetti}, {Bakanov}, {Bamford},
  {Barentsen}, {Barmby}, {Baumbach}, {Berry}, {Biscani}, {Boquien}, {Bostroem},
  {Bouma}, {Brammer}, {Bray}, {Breytenbach}, {Buddelmeijer}, {Burke},
  {Calderone}, {Cano Rodr{\'\i}guez}, {Cara}, {Cardoso}, {Cheedella}, {Copin},
  {Corrales}, {Crichton}, {D'Avella}, {Deil}, {Depagne}, {Dietrich}, {Donath},
  {Droettboom}, {Earl}, {Erben}, {Fabbro}, {Ferreira}, {Finethy}, {Fox},
  {Garrison}, {Gibbons}, {Goldstein}, {Gommers}, {Greco}, {Greenfield},
  {Groener}, {Grollier}, {Hagen}, {Hirst}, {Homeier}, {Horton}, {Hosseinzadeh},
  {Hu}, {Hunkeler}, {Ivezi{\'c}}, {Jain}, {Jenness}, {Kanarek}, {Kendrew},
  {Kern}, {Kerzendorf}, {Khvalko}, {King}, {Kirkby}, {Kulkarni}, {Kumar},
  {Lee}, {Lenz}, {Littlefair}, {Ma}, {Macleod}, {Mastropietro}, {McCully},
  {Montagnac}, {Morris}, {Mueller}, {Mumford}, {Muna}, {Murphy}, {Nelson},
  {Nguyen}, {Ninan}, {N{\"o}the}, {Ogaz}, {Oh}, {Parejko}, {Parley}, {Pascual},
  {Patil}, {Patil}, {Plunkett}, {Prochaska}, {Rastogi}, {Reddy Janga},
  {Sabater}, {Sakurikar}, {Seifert}, {Sherbert}, {Sherwood-Taylor}, {Shih},
  {Sick}, {Silbiger}, {Singanamalla}, {Singer}, {Sladen}, {Sooley},
  {Sornarajah}, {Streicher}, {Teuben}, {Thomas}, {Tremblay}, {Turner},
  {Terr{\'o}n}, {van Kerkwijk}, {de la Vega}, {Watkins}, {Weaver}, {Whitmore},
  {Woillez}, {Zabalza}, \& {Astropy Contributors}}]{2018AJ....156..123A}
{Astropy Collaboration}, {Price-Whelan}, A.~M., {Sip{\H{o}}cz}, B.~M., {et~al.}
  2018, \aj, 156, 123

\bibitem[{{Atri} \& {Mogan}(2021)}]{2021MNRAS.500L...1A}
{Atri}, D., \& {Mogan}, S. R.~C. 2021, \mnras, 500, L1

\bibitem[{{Bai} {et~al.}(2023){Bai}, {Wang}, {Li}, {Xin}, {Li}, {Yang}, \&
  {Wei}}]{2023PASP..135f4201B}
{Bai}, J.-Y., {Wang}, J., {Li}, H.~L., {et~al.} 2023, \pasp, 135, 064201

\bibitem[{{Berger} {et~al.}(2010){Berger}, {Basri}, {Fleming}, {Giampapa},
  {Gizis}, {Liebert}, {Mart{\'\i}n}, {Phan-Bao}, \&
  {Rutledge}}]{2010ApJ...709..332B}
{Berger}, E., {Basri}, G., {Fleming}, T.~A., {et~al.} 2010, \apj, 709, 332

\bibitem[{{Bonnarel} {et~al.}(2000){Bonnarel}, {Fernique}, {Bienaym{\'e}},
  {Egret}, {Genova}, {Louys}, {Ochsenbein}, {Wenger}, \&
  {Bartlett}}]{2000A&AS..143...33B}
{Bonnarel}, F., {Fernique}, P., {Bienaym{\'e}}, O., {et~al.} 2000, \aaps, 143,
  33

\bibitem[{{Boteler}(2006)}]{2006AdSpR..38..159B}
{Boteler}, D.~H. 2006, Advances in Space Research, 38, 159

\bibitem[{{Boudreaux} {et~al.}(2022){Boudreaux}, {Newton}, {Mondrik},
  {Charbonneau}, \& {Irwin}}]{2022ApJ...929...80B}
{Boudreaux}, E.~M., {Newton}, E.~R., {Mondrik}, N., {Charbonneau}, D., \&
  {Irwin}, J. 2022, \apj, 929, 80

\bibitem[{{Carrington}(1859)}]{1859MNRAS..20...13C}
{Carrington}, R.~C. 1859, \mnras, 20, 13

\bibitem[{{Chambers} {et~al.}(2016){Chambers}, {Magnier}, {Metcalfe},
  {Flewelling}, {Huber}, {Waters}, {Denneau}, {Draper}, {Farrow}, {Finkbeiner},
  {Holmberg}, {Koppenhoefer}, {Price}, {Rest}, {Saglia}, {Schlafly}, {Smartt},
  {Sweeney}, {Wainscoat}, {Burgett}, {Chastel}, {Grav}, {Heasley}, {Hodapp},
  {Jedicke}, {Kaiser}, {Kudritzki}, {Luppino}, {Lupton}, {Monet}, {Morgan},
  {Onaka}, {Shiao}, {Stubbs}, {Tonry}, {White}, {Ba{\~n}ados}, {Bell},
  {Bender}, {Bernard}, {Boegner}, {Boffi}, {Botticella}, {Calamida},
  {Casertano}, {Chen}, {Chen}, {Cole}, {Deacon}, {Frenk}, {Fitzsimmons},
  {Gezari}, {Gibbs}, {Goessl}, {Goggia}, {Gourgue}, {Goldman}, {Grant},
  {Grebel}, {Hambly}, {Hasinger}, {Heavens}, {Heckman}, {Henderson}, {Henning},
  {Holman}, {Hopp}, {Ip}, {Isani}, {Jackson}, {Keyes}, {Koekemoer}, {Kotak},
  {Le}, {Liska}, {Long}, {Lucey}, {Liu}, {Martin}, {Masci}, {McLean}, {Mindel},
  {Misra}, {Morganson}, {Murphy}, {Obaika}, {Narayan}, {Nieto-Santisteban},
  {Norberg}, {Peacock}, {Pier}, {Postman}, {Primak}, {Rae}, {Rai}, {Riess},
  {Riffeser}, {Rix}, {R{\"o}ser}, {Russel}, {Rutz}, {Schilbach}, {Schultz},
  {Scolnic}, {Strolger}, {Szalay}, {Seitz}, {Small}, {Smith}, {Soderblom},
  {Taylor}, {Thomson}, {Taylor}, {Thakar}, {Thiel}, {Thilker}, {Unger},
  {Urata}, {Valenti}, {Wagner}, {Walder}, {Walter}, {Watters}, {Werner},
  {Wood-Vasey}, \& {Wyse}}]{2016arXiv161205560C}
{Chambers}, K.~C., {Magnier}, E.~A., {Metcalfe}, N., {et~al.} 2016, arXiv
  e-prints, arXiv:1612.05560

\bibitem[{{Chen} {et~al.}(2021){Chen}, {Zhan}, {Youngblood}, {Wolf},
  {Feinstein}, \& {Horton}}]{2021NatAs...5..298C}
{Chen}, H., {Zhan}, Z., {Youngblood}, A., {et~al.} 2021, Nature Astronomy, 5,
  298

\bibitem[{{Chen} {et~al.}(2014){Chen}, {Girardi}, {Bressan}, {Marigo},
  {Barbieri}, \& {Kong}}]{2014MNRAS.444.2525C}
{Chen}, Y., {Girardi}, L., {Bressan}, A., {et~al.} 2014, \mnras, 444, 2525

\bibitem[{{Cifuentes} {et~al.}(2020){Cifuentes}, {Caballero},
  {Cort{\'e}s-Contreras}, {Montes}, {Abell{\'a}n}, {Dorda}, {Holgado},
  {Zapatero Osorio}, {Morales}, {Amado}, {Passegger}, {Quirrenbach}, {Reiners},
  {Ribas}, {Sanz-Forcada}, {Schweitzer}, {Seifert}, \&
  {Solano}}]{2020A&A...642A.115C}
{Cifuentes}, C., {Caballero}, J.~A., {Cort{\'e}s-Contreras}, M., {et~al.} 2020,
  \aap, 642, A115

\bibitem[{{Cui} {et~al.}(2012){Cui}, {Zhao}, {Chu}, {Li}, {Li}, {Zhang}, {Su},
  {Yao}, {Wang}, {Xing}, {Li}, {Zhu}, {Wang}, {Gu}, {Luo}, {Xu}, {Zhang},
  {Liu}, {Zhang}, {Yang}, {Cao}, {Chen}, {Chen}, {Chen}, {Chen}, {Chu}, {Feng},
  {Gong}, {Hou}, {Hu}, {Hu}, {Hu}, {Jia}, {Jiang}, {Jiang}, {Jiang}, {Jin},
  {Li}, {Li}, {Li}, {Liu}, {Liu}, {Lu}, {Mao}, {Men}, {Qi}, {Qi}, {Shi},
  {Tang}, {Tao}, {Wang}, {Wang}, {Wang}, {Wang}, {Wang}, {Wang}, {Wang},
  {Wang}, {Wang}, {Wang}, {Wang}, {Wang}, {Xu}, {Xu}, {Yang}, {Yu}, {Yuan},
  {Yuan}, {Zhai}, {Zhang}, {Zhang}, {Zhang}, {Zhao}, {Zhou}, {Zhou}, {Zhu}, \&
  {Zou}}]{2012RAA....12.1197C}
{Cui}, X.-Q., {Zhao}, Y.-H., {Chu}, Y.-Q., {et~al.} 2012, Research in Astronomy
  and Astrophysics, 12, 1197

\bibitem[{{Deng} {et~al.}(2012){Deng}, {Newberg}, {Liu}, {Carlin}, {Beers},
  {Chen}, {Chen}, {Christlieb}, {Grillmair}, {Guhathakurta}, {Han}, {Hou},
  {Lee}, {L{\'e}pine}, {Li}, {Liu}, {Pan}, {Sellwood}, {Wang}, {Wang}, {Yang},
  {Yanny}, {Zhang}, {Zhang}, {Zheng}, \& {Zhu}}]{2012RAA....12..735D}
{Deng}, L.-C., {Newberg}, H.~J., {Liu}, C., {et~al.} 2012, Research in
  Astronomy and Astrophysics, 12, 735

\bibitem[{{Esplin} \& {Luhman}(2019)}]{2019AJ....158...54E}
{Esplin}, T.~L., \& {Luhman}, K.~L. 2019, \aj, 158, 54

\bibitem[{{Estrela} \& {Valio}(2018)}]{2018AsBio..18.1414E}
{Estrela}, R., \& {Valio}, A. 2018, Astrobiology, 18, 1414

\bibitem[{{Gaia Collaboration}(2022)}]{2022yCat.1355....0G}
{Gaia Collaboration}. 2022, VizieR Online Data Catalog, I/355

\bibitem[{{Gaia Collaboration} {et~al.}(2018){Gaia Collaboration}, {Babusiaux},
  {van Leeuwen}, {Barstow}, {Jordi}, {Vallenari}, {Bossini}, {Bressan},
  {Cantat-Gaudin}, {van Leeuwen}, {Brown}, {Prusti}, {de Bruijne},
  {Bailer-Jones}, {Biermann}, {Evans}, {Eyer}, {Jansen}, {Klioner}, {Lammers},
  {Lindegren}, {Luri}, {Mignard}, {Panem}, {Pourbaix}, {Randich}, {Sartoretti},
  {Siddiqui}, {Soubiran}, {Walton}, {Arenou}, {Bastian}, {Cropper}, {Drimmel},
  {Katz}, {Lattanzi}, {Bakker}, {Cacciari}, {Casta{\~n}eda}, {Chaoul}, {Cheek},
  {De Angeli}, {Fabricius}, {Guerra}, {Holl}, {Masana}, {Messineo}, {Mowlavi},
  {Nienartowicz}, {Panuzzo}, {Portell}, {Riello}, {Seabroke}, {Tanga},
  {Th{\'e}venin}, {Gracia-Abril}, {Comoretto}, {Garcia-Reinaldos}, {Teyssier},
  {Altmann}, {Andrae}, {Audard}, {Bellas-Velidis}, {Benson}, {Berthier},
  {Blomme}, {Burgess}, {Busso}, {Carry}, {Cellino}, {Clementini}, {Clotet},
  {Creevey}, {Davidson}, {De Ridder}, {Delchambre}, {Dell'Oro}, {Ducourant},
  {Fern{\'a}ndez-Hern{\'a}ndez}, {Fouesneau}, {Fr{\'e}mat}, {Galluccio},
  {Garc{\'\i}a-Torres}, {Gonz{\'a}lez-N{\'u}{\~n}ez}, {Gonz{\'a}lez-Vidal},
  {Gosset}, {Guy}, {Halbwachs}, {Hambly}, {Harrison}, {Hern{\'a}ndez},
  {Hestroffer}, {Hodgkin}, {Hutton}, {Jasniewicz}, {Jean-Antoine-Piccolo},
  {Jordan}, {Korn}, {Krone-Martins}, {Lanzafame}, {Lebzelter}, {L{\"o}ffler},
  {Manteiga}, {Marrese}, {Mart{\'\i}n-Fleitas}, {Moitinho}, {Mora}, {Muinonen},
  {Osinde}, {Pancino}, {Pauwels}, {Petit}, {Recio-Blanco}, {Richards},
  {Rimoldini}, {Robin}, {Sarro}, {Siopis}, {Smith}, {Sozzetti}, {S{\"u}veges},
  {Torra}, {van Reeven}, {Abbas}, {Abreu Aramburu}, {Accart}, {Aerts},
  {Altavilla}, {{\'A}lvarez}, {Alvarez}, {Alves}, {Anderson}, {Andrei},
  {Anglada Varela}, {Antiche}, {Antoja}, {Arcay}, {Astraatmadja}, {Bach},
  {Baker}, {Balaguer-N{\'u}{\~n}ez}, {Balm}, {Barache}, {Barata}, {Barbato},
  {Barblan}, {Barklem}, {Barrado}, {Barros}, {Bartholom{\'e} Mu{\~n}oz},
  {Bassilana}, {Becciani}, {Bellazzini}, {Berihuete}, {Bertone}, {Bianchi},
  {Bienaym{\'e}}, {Blanco-Cuaresma}, {Boch}, {Boeche}, {Bombrun}, {Borrachero},
  {Bouquillon}, {Bourda}, {Bragaglia}, {Bramante}, {Breddels}, {Brouillet},
  {Br{\"u}semeister}, {Brugaletta}, {Bucciarelli}, {Burlacu}, {Busonero},
  {Butkevich}, {Buzzi}, {Caffau}, {Cancelliere}, {Cannizzaro}, {Carballo},
  {Carlucci}, {Carrasco}, {Casamiquela}, {Castellani}, {Castro-Ginard},
  {Charlot}, {Chemin}, {Chiavassa}, {Cocozza}, {Costigan}, {Cowell}, {Crifo},
  {Crosta}, {Crowley}, {Cuypers}, {Dafonte}, {Damerdji}, {Dapergolas}, {David},
  {David}, {de Laverny}, {De Luise}, {De March}, {de Martino}, {de Souza}, {de
  Torres}, {Debosscher}, {del Pozo}, {Delbo}, {Delgado}, {Delgado}, {Diakite},
  {Diener}, {Distefano}, {Dolding}, {Drazinos}, {Dur{\'a}n}, {Edvardsson},
  {Enke}, {Eriksson}, {Esquej}, {Eynard Bontemps}, {Fabre}, {Fabrizio},
  {Faigler}, {Falc{\~a}o}, {Farr{\`a}s Casas}, {Federici}, {Fedorets},
  {Fernique}, {Figueras}, {Filippi}, {Findeisen}, {Fonti}, {Fraile}, {Fraser},
  {Fr{\'e}zouls}, {Gai}, {Galleti}, {Garabato}, {Garc{\'\i}a-Sedano},
  {Garofalo}, {Garralda}, {Gavel}, {Gavras}, {Gerssen}, {Geyer}, {Giacobbe},
  {Gilmore}, {Girona}, {Giuffrida}, {Glass}, {Gomes}, {Granvik}, {Gueguen},
  {Guerrier}, {Guiraud}, {Guti{\'e}}, {Haigron}, {Hatzidimitriou}, {Hauser},
  {Haywood}, {Heiter}, {Helmi}, {Heu}, {Hilger}, {Hobbs}, {Hofmann}, {Holland},
  {Huckle}, {Hypki}, {Icardi}, {Jan{\ss}en}, {Jevardat de Fombelle}, {Jonker},
  {Juh{\'a}sz}, {Julbe}, {Karampelas}, {Kewley}, {Klar}, {Kochoska}, {Kohley},
  {Kolenberg}, {Kontizas}, {Kontizas}, {Koposov}, {Kordopatis},
  {Kostrzewa-Rutkowska}, {Koubsky}, {Lambert}, {Lanza}, {Lasne}, {Lavigne}, {Le
  Fustec}, {Le Poncin-Lafitte}, {Lebreton}, {Leccia}, {Leclerc},
  {Lecoeur-Taibi}, {Lenhardt}, {Leroux}, {Liao}, {Licata}, {Lindstr{\o}m},
  {Lister}, {Livanou}, {Lobel}, {L{\'o}pez}, {Managau}, {Mann}, {Mantelet},
  {Marchal}, {Marchant}, {Marconi}, {Marinoni}, {Marschalk{\'o}}, {Marshall},
  {Martino}, {Marton}, {Mary}, {Massari}, {Matijevi{\v{c}}}, {Mazeh},
  {McMillan}, {Messina}, {Michalik}, {Millar}, {Molina}, {Molinaro},
  {Moln{\'a}r}, {Montegriffo}, {Mor}, {Morbidelli}, {Morel}, {Morris},
  {Mulone}, {Muraveva}, {Musella}, {Nelemans}, {Nicastro}, {Noval},
  {O'Mullane}, {Ord{\'e}novic}, {Ord{\'o}{\~n}ez-Blanco}, {Osborne}, {Pagani},
  {Pagano}, {Pailler}, {Palacin}, {Palaversa}, {Panahi}, {Pawlak},
  {Piersimoni}, {Pineau}, {Plachy}, {Plum}, {Poggio}, {Poujoulet},
  {Pr{\v{s}}a}, {Pulone}, {Racero}, {Ragaini}, {Rambaux}, {Ramos-Lerate},
  {Regibo}, {Reyl{\'e}}, {Riclet}, {Ripepi}, {Riva}, {Rivard}, {Rixon},
  {Roegiers}, {Roelens}, {Romero-G{\'o}mez}, {Rowell}, {Royer}, {Ruiz-Dern},
  {Sadowski}, {Sagrist{\`a} Sell{\'e}s}, {Sahlmann}, {Salgado}, {Salguero},
  {Sanna}, {Santana-Ros}, {Sarasso}, {Savietto}, {Schultheis}, {Sciacca},
  {Segol}, {Segovia}, {S{\'e}gransan}, {Shih}, {Siltala}, {Silva}, {Smart},
  {Smith}, {Solano}, {Solitro}, {Sordo}, {Soria Nieto}, {Souchay}, {Spagna},
  {Spoto}, {Stampa}, {Steele}, {Steidelm{\"u}ller}, {Stephenson}, {Stoev},
  {Suess}, {Surdej}, {Szabados}, {Szegedi-Elek}, {Tapiador}, {Taris}, {Tauran},
  {Taylor}, {Teixeira}, {Terrett}, {Teyssandier}, {Thuillot}, {Titarenko},
  {Torra Clotet}, {Turon}, {Ulla}, {Utrilla}, {Uzzi}, {Vaillant}, {Valentini},
  {Valette}, {van Elteren}, {Van Hemelryck}, {Vaschetto}, {Vecchiato},
  {Veljanoski}, {Viala}, {Vicente}, {Vogt}, {von Essen}, {Voss}, {Votruba},
  {Voutsinas}, {Walmsley}, {Weiler}, {Wertz}, {Wevers}, {Wyrzykowski},
  {Yoldas}, {{\v{Z}}erjal}, {Ziaeepour}, {Zorec}, {Zschocke}, {Zucker},
  {Zurbach}, \& {Zwitter}}]{2018A&A...616A..10G}
{Gaia Collaboration}, {Babusiaux}, C., {van Leeuwen}, F., {et~al.} 2018, \aap,
  616, A10

\bibitem[{{Gershberg}(1972)}]{1972Ap&SS..19...75G}
{Gershberg}, R.~E. 1972, \apss, 19, 75

\bibitem[{{G{\"u}nther} {et~al.}(2020){G{\"u}nther}, {Zhan}, {Seager},
  {Rimmer}, {Ranjan}, {Stassun}, {Oelkers}, {Daylan}, {Newton}, {Kristiansen},
  {Olah}, {Gillen}, {Rappaport}, {Ricker}, {Vanderspek}, {Latham}, {Winn},
  {Jenkins}, {Glidden}, {Fausnaugh}, {Levine}, {Dittmann}, {Quinn},
  {Krishnamurthy}, \& {Ting}}]{2020AJ....159...60G}
{G{\"u}nther}, M.~N., {Zhan}, Z., {Seager}, S., {et~al.} 2020, \aj, 159, 60

\bibitem[{{Haisch} {et~al.}(1991){Haisch}, {Strong}, \&
  {Rodono}}]{1991ARA&A..29..275H}
{Haisch}, B., {Strong}, K.~T., \& {Rodono}, M. 1991, \araa, 29, 275

\bibitem[{{Han} {et~al.}(2021){Han}, {Xiao}, {Zhang}, {Turpin}, {Xin}, {Wu},
  {Cai}, {Dong}, {Huang}, {Kang}, {Leroy}, {Li}, {Li}, {Lu}, {Qiu}, {Stahl},
  {Wang}, {Wang}, {Xu}, {Yang}, {Zhao}, {Zhang}, {Zheng}, {Zheng}, \&
  {Wei}}]{2021PASP..133f5001H}
{Han}, X., {Xiao}, Y., {Zhang}, P., {et~al.} 2021, \pasp, 133, 065001

\bibitem[{{Hodgson}(1859)}]{1859MNRAS..20...15H}
{Hodgson}, R. 1859, \mnras, 20, 15

\bibitem[{{Howard} {et~al.}(2019){Howard}, {Corbett}, {Law}, {Ratzloff},
  {Glazier}, {Fors}, {del Ser}, \& {Haislip}}]{2019ApJ...881....9H}
{Howard}, W.~S., {Corbett}, H., {Law}, N.~M., {et~al.} 2019, \apj, 881, 9

\bibitem[{{Howard} \& {MacGregor}(2022)}]{2022ApJ...926..204H}
{Howard}, W.~S., \& {MacGregor}, M.~A. 2022, \apj, 926, 204

\bibitem[{{Howard} {et~al.}(2020){Howard}, {Corbett}, {Law}, {Ratzloff},
  {Galliher}, {Glazier}, {Gonzalez}, {Vasquez Soto}, {Fors}, {del Ser}, \&
  {Haislip}}]{2020ApJ...902..115H}
{Howard}, W.~S., {Corbett}, H., {Law}, N.~M., {et~al.} 2020, \apj, 902, 115

\bibitem[{{Hu} {et~al.}(2022){Hu}, {Airapetian}, {Li}, {Zank}, \&
  {Jin}}]{2022SciA....8I9743H}
{Hu}, J., {Airapetian}, V.~S., {Li}, G., {Zank}, G., \& {Jin}, M. 2022, Science
  Advances, 8, eabi9743

\bibitem[{{Huber} {et~al.}(2016){Huber}, {Bryson}, {Haas}, {Barclay},
  {Barentsen}, {Howell}, {Sharma}, {Stello}, \&
  {Thompson}}]{2016ApJS..224....2H}
{Huber}, D., {Bryson}, S.~T., {Haas}, M.~R., {et~al.} 2016, \apjs, 224, 2

\bibitem[{{Hudson}(2021)}]{2021ARA&A..59..445H}
{Hudson}, H.~S. 2021, \araa, 59, 445

\bibitem[{{Jackman} {et~al.}(2023){Jackman}, {Wheatley}, {West}, {Gill}, \&
  {Jenkins}}]{2023MNRAS.525.1588J}
{Jackman}, J. A.~G., {Wheatley}, P.~J., {West}, R.~G., {Gill}, S., \&
  {Jenkins}, J.~S. 2023, \mnras, 525, 1588

\bibitem[{{Jackman} {et~al.}(2019){Jackman}, {Wheatley}, {Bayliss}, {Burleigh},
  {Casewell}, {Eigm{\"u}ller}, {Goad}, {Pollacco}, {Raynard}, {Watson}, \&
  {West}}]{2019MNRAS.485L.136J}
{Jackman}, J. A.~G., {Wheatley}, P.~J., {Bayliss}, D., {et~al.} 2019, \mnras,
  485, L136

\bibitem[{{Jackman} {et~al.}(2021){Jackman}, {Wheatley}, {Acton}, {Anderson},
  {Bayliss}, {Briegal}, {Burleigh}, {Casewell}, {G{\"a}nsicke}, {Gill},
  {Gillen}, {Goad}, {G{\"u}nther}, {Henderson}, {Hodgkin}, {Jenkins}, {Pugh},
  {Queloz}, {Raynard}, {Tilbrook}, {Watson}, \& {West}}]{2021MNRAS.504.3246J}
{Jackman}, J. A.~G., {Wheatley}, P.~J., {Acton}, J.~S., {et~al.} 2021, \mnras,
  504, 3246

\bibitem[{{Jao} {et~al.}(2022){Jao}, {Couperus}, {Vrijmoet}, {Wright}, \&
  {Henry}}]{2022ApJ...940..145J}
{Jao}, W.-C., {Couperus}, A.~A., {Vrijmoet}, E.~H., {Wright}, N.~J., \&
  {Henry}, T.~J. 2022, \apj, 940, 145

\bibitem[{{Kane}(2018)}]{2018ApJ...861L..21K}
{Kane}, S.~R. 2018, \apjl, 861, L21

\bibitem[{{Kiman} {et~al.}(2021){Kiman}, {Faherty}, {Cruz}, {Gagn{\'e}},
  {Angus}, {Schmidt}, {Mann}, {Bardalez Gagliuffi}, \&
  {Rice}}]{2021AJ....161..277K}
{Kiman}, R., {Faherty}, J.~K., {Cruz}, K.~L., {et~al.} 2021, \aj, 161, 277

\bibitem[{{Kirkpatrick} {et~al.}(1991){Kirkpatrick}, {Henry}, \&
  {McCarthy}}]{1991ApJS...77..417K}
{Kirkpatrick}, J.~D., {Henry}, T.~J., \& {McCarthy}, Donald~W., J. 1991, \apjs,
  77, 417

\bibitem[{{Konings} {et~al.}(2022){Konings}, {Baeyens}, \&
  {Decin}}]{2022A&A...667A..15K}
{Konings}, T., {Baeyens}, R., \& {Decin}, L. 2022, \aap, 667, A15

\bibitem[{{Kowalski} {et~al.}(2013){Kowalski}, {Hawley}, {Wisniewski}, {Osten},
  {Hilton}, {Holtzman}, {Schmidt}, \& {Davenport}}]{2013ApJS..207...15K}
{Kowalski}, A.~F., {Hawley}, S.~L., {Wisniewski}, J.~P., {et~al.} 2013, \apjs,
  207, 15

\bibitem[{{Lehtinen} {et~al.}(2021){Lehtinen}, {K{\"a}pyl{\"a}}, {Olspert}, \&
  {Spada}}]{2021ApJ...910..110L}
{Lehtinen}, J.~J., {K{\"a}pyl{\"a}}, M.~J., {Olspert}, N., \& {Spada}, F. 2021,
  \apj, 910, 110

\bibitem[{{Li} {et~al.}(2023{\natexlab{a}}){Li}, {Wu}, {Zhou}, {Yang}, {Li},
  {Chen}, {Xin}, {Wang}, {Haerken}, {Ma}, {Cai}, {Han}, {Huang}, {Lu}, {Bai},
  {Zhang}, {Hao}, {Wang}, {Dai}, {Liang}, {Meng}, \&
  {Wei}}]{2023RAA....23a5016L}
{Li}, G.-W., {Wu}, C., {Zhou}, G.-P., {et~al.} 2023{\natexlab{a}}, Research in
  Astronomy and Astrophysics, 23, 015016

\bibitem[{{Li} {et~al.}(2023{\natexlab{b}}){Li}, {Wang}, {Xin}, {Bai}, {Han},
  {Cai}, {Huang}, {Lu}, {Qiu}, {Wu}, {Li}, {Deng}, {Xu}, {Yang}, {Wang},
  {Liang}, \& {Wei}}]{2023ApJ...954..142L}
{Li}, H.-L., {Wang}, J., {Xin}, L.-P., {et~al.} 2023{\natexlab{b}}, \apj, 954,
  142

\bibitem[{{Li} {et~al.}(2021){Li}, {Chen}, {Hou}, {Veronig}, {Yang}, \&
  {Zhang}}]{2021ApJ...917L..29L}
{Li}, T., {Chen}, A., {Hou}, Y., {et~al.} 2021, \apjl, 917, L29

\bibitem[{{Lightkurve Collaboration} {et~al.}(2018){Lightkurve Collaboration},
  {Cardoso}, {Hedges}, {Gully-Santiago}, {Saunders}, {Cody}, {Barclay}, {Hall},
  {Sagear}, {Turtelboom}, {Zhang}, {Tzanidakis}, {Mighell}, {Coughlin}, {Bell},
  {Berta-Thompson}, {Williams}, {Dotson}, \& {Barentsen}}]{2018ascl.soft12013L}
{Lightkurve Collaboration}, {Cardoso}, J.~V.~d.~M., {Hedges}, C., {et~al.}
  2018, {Lightkurve: Kepler and TESS time series analysis in Python},
  Astrophysics Source Code Library, , , ascl:1812.013

\bibitem[{{Linsky}(2019)}]{2019LNP...955.....L}
{Linsky}, J. 2019, {Host Stars and their Effects on Exoplanet Atmospheres},
  Vol. 955, doi:10.1007/978-3-030-11452-7

\bibitem[{{Lomb}(1976)}]{1976Ap&SS..39..447L}
{Lomb}, N.~R. 1976, \apss, 39, 447

\bibitem[{{Mamajek} {et~al.}(2015){Mamajek}, {Torres}, {Prsa}, {Harmanec},
  {Asplund}, {Bennett}, {Capitaine}, {Christensen-Dalsgaard}, {Depagne},
  {Folkner}, {Haberreiter}, {Hekker}, {Hilton}, {Kostov}, {Kurtz}, {Laskar},
  {Mason}, {Milone}, {Montgomery}, {Richards}, {Schou}, \&
  {Stewart}}]{2015arXiv151006262M}
{Mamajek}, E.~E., {Torres}, G., {Prsa}, A., {et~al.} 2015, arXiv e-prints,
  arXiv:1510.06262

\bibitem[{{Medina} {et~al.}(2022){Medina}, {Winters}, {Irwin}, \&
  {Charbonneau}}]{2022ApJ...935..104M}
{Medina}, A.~A., {Winters}, J.~G., {Irwin}, J.~M., \& {Charbonneau}, D. 2022,
  \apj, 935, 104

\bibitem[{{Miranda-Rosete} {et~al.}(2023){Miranda-Rosete}, {Segura}, \&
  {Schwieterman}}]{2023RMxAC..55...99M}
{Miranda-Rosete}, A., {Segura}, A., \& {Schwieterman}, E.~W. 2023, in Revista
  Mexicana de Astronomia y Astrofisica Conference Series, Vol.~55, Revista
  Mexicana de Astronomia y Astrofisica Conference Series, 99--99

\bibitem[{{Mohanty} \& {Basri}(2003)}]{2003ApJ...583..451M}
{Mohanty}, S., \& {Basri}, G. 2003, \apj, 583, 451

\bibitem[{{Mohanty} {et~al.}(2002){Mohanty}, {Basri}, {Shu}, {Allard}, \&
  {Chabrier}}]{2002ApJ...571..469M}
{Mohanty}, S., {Basri}, G., {Shu}, F., {Allard}, F., \& {Chabrier}, G. 2002,
  \apj, 571, 469

\bibitem[{{Moreno C{\'a}rdenas} {et~al.}(2016){Moreno C{\'a}rdenas},
  {Cristancho S{\'a}nchez}, \& {Vargas Dom{\'\i}nguez}}]{2016AdSpR..57..257M}
{Moreno C{\'a}rdenas}, F., {Cristancho S{\'a}nchez}, S., \& {Vargas
  Dom{\'\i}nguez}, S. 2016, Advances in Space Research, 57, 257

\bibitem[{{Newton} {et~al.}(2017){Newton}, {Irwin}, {Charbonneau}, {Berlind},
  {Calkins}, \& {Mink}}]{2017ApJ...834...85N}
{Newton}, E.~R., {Irwin}, J., {Charbonneau}, D., {et~al.} 2017, \apj, 834, 85

\bibitem[{{Notsu} {et~al.}(2019){Notsu}, {Maehara}, {Honda}, {Hawley},
  {Davenport}, {Namekata}, {Notsu}, {Ikuta}, {Nogami}, \&
  {Shibata}}]{2019ApJ...876...58N}
{Notsu}, Y., {Maehara}, H., {Honda}, S., {et~al.} 2019, \apj, 876, 58

\bibitem[{{Osten} {et~al.}(2005){Osten}, {Hawley}, {Allred}, {Johns-Krull}, \&
  {Roark}}]{2005ApJ...621..398O}
{Osten}, R.~A., {Hawley}, S.~L., {Allred}, J.~C., {Johns-Krull}, C.~M., \&
  {Roark}, C. 2005, \apj, 621, 398

\bibitem[{{Paegert} {et~al.}(2022){Paegert}, {Stassun}, {Collins}, {Pepper},
  {Torres}, {Jenkins}, {Twicken}, \& {Latham}}]{2022yCat.4039....0P}
{Paegert}, M., {Stassun}, K.~G., {Collins}, K.~A., {et~al.} 2022, VizieR Online
  Data Catalog, IV/39

\bibitem[{{Pass} {et~al.}(2022){Pass}, {Charbonneau}, {Irwin}, \&
  {Winters}}]{2022ApJ...936..109P}
{Pass}, E.~K., {Charbonneau}, D., {Irwin}, J.~M., \& {Winters}, J.~G. 2022,
  \apj, 936, 109

\bibitem[{{Pietras} {et~al.}(2022){Pietras}, {Falewicz}, {Siarkowski}, {Bicz},
  \& {Pre{\'s}}}]{2022ApJ...935..143P}
{Pietras}, M., {Falewicz}, R., {Siarkowski}, M., {Bicz}, K., \& {Pre{\'s}}, P.
  2022, \apj, 935, 143

\bibitem[{{Popinchalk} {et~al.}(2021){Popinchalk}, {Faherty}, {Kiman},
  {Gagn{\'e}}, {Curtis}, {Angus}, {Cruz}, \& {Rice}}]{2021ApJ...916...77P}
{Popinchalk}, M., {Faherty}, J.~K., {Kiman}, R., {et~al.} 2021, \apj, 916, 77

\bibitem[{{Raetz} {et~al.}(2020){Raetz}, {Stelzer}, {Damasso}, \&
  {Scholz}}]{2020A&A...637A..22R}
{Raetz}, S., {Stelzer}, B., {Damasso}, M., \& {Scholz}, A. 2020, \aap, 637, A22

\bibitem[{{Rebassa-Mansergas} {et~al.}(2012){Rebassa-Mansergas}, {Nebot
  G{\'o}mez-Mor{\'a}n}, {Schreiber}, {G{\"a}nsicke}, {Schwope}, {Gallardo}, \&
  {Koester}}]{2012MNRAS.419..806R}
{Rebassa-Mansergas}, A., {Nebot G{\'o}mez-Mor{\'a}n}, A., {Schreiber}, M.~R.,
  {et~al.} 2012, \mnras, 419, 806

\bibitem[{{Reiners} {et~al.}(2014){Reiners}, {Sch{\"u}ssler}, \&
  {Passegger}}]{2014ApJ...794..144R}
{Reiners}, A., {Sch{\"u}ssler}, M., \& {Passegger}, V.~M. 2014, \apj, 794, 144

\bibitem[{{Rimmer} {et~al.}(2018){Rimmer}, {Xu}, {Thompson}, {Gillen},
  {Sutherland}, \& {Queloz}}]{2018SciA....4.3302R}
{Rimmer}, P.~B., {Xu}, J., {Thompson}, S.~J., {et~al.} 2018, Science Advances,
  4, eaar3302

\bibitem[{{Scargle}(1982)}]{1982ApJ...263..835S}
{Scargle}, J.~D. 1982, \apj, 263, 835

\bibitem[{{Schmidt} {et~al.}(2015){Schmidt}, {Hawley}, {West}, {Bochanski},
  {Davenport}, {Ge}, \& {Schneider}}]{2015AJ....149..158S}
{Schmidt}, S.~J., {Hawley}, S.~L., {West}, A.~A., {et~al.} 2015, \aj, 149, 158

\bibitem[{{Schmidt} {et~al.}(2016){Schmidt}, {Shappee}, {Gagn{\'e}}, {Stanek},
  {Prieto}, {Holoien}, {Kochanek}, {Chomiuk}, {Dong}, {Seibert}, \&
  {Strader}}]{2016ApJ...828L..22S}
{Schmidt}, S.~J., {Shappee}, B.~J., {Gagn{\'e}}, J., {et~al.} 2016, \apjl, 828,
  L22

\bibitem[{{Schmidt} {et~al.}(2019){Schmidt}, {Shappee}, {van Saders}, {Stanek},
  {Brown}, {Kochanek}, {Dong}, {Drout}, {Frank}, {Holoien}, {Johnson},
  {Madore}, {Prieto}, {Seibert}, {Seidel}, \& {Simonian}}]{2019ApJ...876..115S}
{Schmidt}, S.~J., {Shappee}, B.~J., {van Saders}, J.~L., {et~al.} 2019, \apj,
  876, 115

\bibitem[{{Sch{\"o}nrich} {et~al.}(2010){Sch{\"o}nrich}, {Binney}, \&
  {Dehnen}}]{2010MNRAS.403.1829S}
{Sch{\"o}nrich}, R., {Binney}, J., \& {Dehnen}, W. 2010, \mnras, 403, 1829

\bibitem[{{Shibata} \& {Magara}(2011)}]{2011LRSP....8....6S}
{Shibata}, K., \& {Magara}, T. 2011, Living Reviews in Solar Physics, 8, 6

\bibitem[{{Shibata} {et~al.}(2013){Shibata}, {Isobe}, {Hillier}, {Choudhuri},
  {Maehara}, {Ishii}, {Shibayama}, {Notsu}, {Notsu}, {Nagao}, {Honda}, \&
  {Nogami}}]{2013PASJ...65...49S}
{Shibata}, K., {Isobe}, H., {Hillier}, A., {et~al.} 2013, \pasj, 65, 49

\bibitem[{{Shibayama} {et~al.}(2013){Shibayama}, {Maehara}, {Notsu}, {Notsu},
  {Nagao}, {Honda}, {Ishii}, {Nogami}, \& {Shibata}}]{2013ApJS..209....5S}
{Shibayama}, T., {Maehara}, H., {Notsu}, S., {et~al.} 2013, \apjs, 209, 5

\bibitem[{{Skrutskie} {et~al.}(2006){Skrutskie}, {Cutri}, {Stiening},
  {Weinberg}, {Schneider}, {Carpenter}, {Beichman}, {Capps}, {Chester},
  {Elias}, {Huchra}, {Liebert}, {Lonsdale}, {Monet}, {Price}, {Seitzer},
  {Jarrett}, {Kirkpatrick}, {Gizis}, {Howard}, {Evans}, {Fowler}, {Fullmer},
  {Hurt}, {Light}, {Kopan}, {Marsh}, {McCallon}, {Tam}, {Van Dyk}, \&
  {Wheelock}}]{2006AJ....131.1163S}
{Skrutskie}, M.~F., {Cutri}, R.~M., {Stiening}, R., {et~al.} 2006, \aj, 131,
  1163

\bibitem[{{STScI}(2013)}]{https://doi.org/10.17909/T9H59D}
{STScI}. 2013, GALEX/MCAT,  STScI/MAST, doi:10.17909/T9H59D

\bibitem[{{STScI}(2016)}]{https://doi.org/10.17909/T93W28}
---. 2016, Kepler/EPIC,  STScI/MAST, doi:10.17909/T93W28

\bibitem[{{STScI}(2018)}]{https://doi.org/10.17909/fwdt-2x66}
---. 2018, TESS Input Catalog and Candidate Target List,  STScI/MAST,
  doi:10.17909/FWDT-2X66

\bibitem[{{STScI}(2022{\natexlab{a}})}]{https://doi.org/10.17909/55e7-5x63}
---. 2022{\natexlab{a}}, Pan-STARRS1 DR1 Catalog,  STScI/MAST,
  doi:10.17909/55E7-5X63

\bibitem[{{STScI}(2022{\natexlab{b}})}]{https://doi.org/10.17909/3y7c-wa45}
---. 2022{\natexlab{b}}, TESS Raw Full Frame Images: All Sectors,  STScI/MAST,
  doi:10.17909/3Y7C-WA45

\bibitem[{{Sullivan} {et~al.}(2015){Sullivan}, {Winn}, {Berta-Thompson},
  {Charbonneau}, {Deming}, {Dressing}, {Latham}, {Levine}, {McCullough},
  {Morton}, {Ricker}, {Vanderspek}, \& {Woods}}]{2015ApJ...809...77S}
{Sullivan}, P.~W., {Winn}, J.~N., {Berta-Thompson}, Z.~K., {et~al.} 2015, \apj,
  809, 77

\bibitem[{{Tilley} {et~al.}(2019){Tilley}, {Segura}, {Meadows}, {Hawley}, \&
  {Davenport}}]{2019AsBio..19...64T}
{Tilley}, M.~A., {Segura}, A., {Meadows}, V., {Hawley}, S., \& {Davenport}, J.
  2019, Astrobiology, 19, 64

\bibitem[{{Wang} {et~al.}(2020){Wang}, {Li}, {Xin}, {Han}, {Meng}, {Brink},
  {Cai}, {Dai}, {Filippenko}, {Hsia}, {Huang}, {Jia}, {Li}, {Li}, {Liang},
  {Lu}, {Mao}, {Qiu}, {Qiu}, {Ren}, {Turpin}, {Wang}, {Wang}, {Wang}, {Wu},
  {Xu}, {Yan}, {Zhang}, {Zheng}, \& {Wei}}]{2020AJ....159...35W}
{Wang}, J., {Li}, H.~L., {Xin}, L.~P., {et~al.} 2020, \aj, 159, 35

\bibitem[{{Wang} {et~al.}(2021){Wang}, {Xin}, {Li}, {Li}, {Sun}, {Gao}, {Han},
  {Dai}, {Liang}, {Wang}, \& {Wei}}]{2021ApJ...916...92W}
{Wang}, J., {Xin}, L.~P., {Li}, H.~L., {et~al.} 2021, \apj, 916, 92

\bibitem[{{Wang} {et~al.}(2022){Wang}, {Li}, {Xin}, {Li}, {Bai}, {Gao}, {Ren},
  {Song}, {Deng}, {Han}, {Dai}, {Liang}, {Wang}, \&
  {Wei}}]{2022ApJ...934...98W}
{Wang}, J., {Li}, H.~L., {Xin}, L.~P., {et~al.} 2022, \apj, 934, 98

\bibitem[{{Wei} {et~al.}(2016){Wei}, {Cordier}, {Antier}, {Antilogus},
  {Atteia}, {Bajat}, {Basa}, {Beckmann}, {Bernardini}, {Boissier}, {Bouchet},
  {Burwitz}, {Claret}, {Dai}, {Daigne}, {Deng}, {Dornic}, {Feng}, {Foglizzo},
  {Gao}, {Gehrels}, {Godet}, {Goldwurm}, {Gonzalez}, {Gosset}, {G{\"o}tz},
  {Gouiffes}, {Grise}, {Gros}, {Guilet}, {Han}, {Huang}, {Huang}, {Jouret},
  {Klotz}, {La Marle}, {Lachaud}, {Le Floch}, {Lee}, {Leroy}, {Li}, {Li}, {Li},
  {Liang}, {Lyu}, {Mercier}, {Migliori}, {Mochkovitch}, {O'Brien}, {Osborne},
  {Paul}, {Perinati}, {Petitjean}, {Piron}, {Qiu}, {Rau}, {Rodriguez},
  {Schanne}, {Tanvir}, {Vangioni}, {Vergani}, {Wang}, {Wang}, {Wang}, {Wang},
  {Watson}, {Webb}, {Wei}, {Willingale}, {Wu}, {Wu}, {Xin}, {Xu}, {Yu}, {Yu},
  {Yu}, {Zhang}, {Zhang}, {Zhang}, \& {Zhou}}]{2016arXiv161006892W}
{Wei}, J., {Cordier}, B., {Antier}, S., {et~al.} 2016, arXiv e-prints,
  arXiv:1610.06892

\bibitem[{{Wright} {et~al.}(2010){Wright}, {Eisenhardt}, {Mainzer}, {Ressler},
  {Cutri}, {Jarrett}, {Kirkpatrick}, {Padgett}, {McMillan}, {Skrutskie},
  {Stanford}, {Cohen}, {Walker}, {Mather}, {Leisawitz}, {Gautier}, {McLean},
  {Benford}, {Lonsdale}, {Blain}, {Mendez}, {Irace}, {Duval}, {Liu}, {Royer},
  {Heinrichsen}, {Howard}, {Shannon}, {Kendall}, {Walsh}, {Larsen}, {Cardon},
  {Schick}, {Schwalm}, {Abid}, {Fabinsky}, {Naes}, \&
  {Tsai}}]{2010AJ....140.1868W}
{Wright}, E.~L., {Eisenhardt}, P. R.~M., {Mainzer}, A.~K., {et~al.} 2010, \aj,
  140, 1868

\bibitem[{{Wright} \& {Drake}(2016)}]{2016Natur.535..526W}
{Wright}, N.~J., \& {Drake}, J.~J. 2016, \nat, 535, 526

\bibitem[{{Wright} {et~al.}(2011){Wright}, {Drake}, {Mamajek}, \&
  {Henry}}]{2011ApJ...743...48W}
{Wright}, N.~J., {Drake}, J.~J., {Mamajek}, E.~E., \& {Henry}, G.~W. 2011,
  \apj, 743, 48

\bibitem[{{Wright} {et~al.}(2018){Wright}, {Newton}, {Williams}, {Drake}, \&
  {Yadav}}]{2018MNRAS.479.2351W}
{Wright}, N.~J., {Newton}, E.~R., {Williams}, P. K.~G., {Drake}, J.~J., \&
  {Yadav}, R.~K. 2018, \mnras, 479, 2351

\bibitem[{{Xin} {et~al.}(2023{\natexlab{a}}){Xin}, {Han}, {Li}, {Zhang},
  {Wang}, {Turpin}, {Yang}, {Qiu}, {Liang}, {Dai}, {Cai}, {Lu}, {Wang},
  {Huang}, {Wang}, {Wu}, {Gao}, {Ren}, {Zhang}, {Yang}, {Deng}, \&
  {Wei}}]{2023NatAs...7..724X}
{Xin}, L., {Han}, X., {Li}, H., {et~al.} 2023{\natexlab{a}}, Nature Astronomy,
  7, 724

\bibitem[{{Xin} {et~al.}(2021){Xin}, {Li}, {Wang}, {Han}, {Xu}, {Meng}, {Cai},
  {Huang}, {Lu}, {Qiu}, {Wang}, {Liang}, {Dai}, {Wang}, {Wu}, {Zhang}, {Li},
  {Turpin}, {Feng}, {Deng}, {Sun}, {Zheng}, {Yang}, \&
  {Wei}}]{2021ApJ...909..106X}
{Xin}, L.~P., {Li}, H.~L., {Wang}, J., {et~al.} 2021, \apj, 909, 106

\bibitem[{{Xin} {et~al.}(2023{\natexlab{b}}){Xin}, {Li}, {Wang}, {Han}, {Cai},
  {Huang}, {Cao}, {Zhu}, {Wang}, {Li}, {Ren}, {Gao}, {Song}, {Huang}, {Lu},
  {Bai}, {Qiu}, {Liang}, {Dai}, {Wang}, {Wu}, {Deng}, {Yang}, \&
  {Wei}}]{2023MNRAS.tmp..951X}
{Xin}, L.-P., {Li}, H.-l., {Wang}, J., {et~al.} 2023{\natexlab{b}}, \mnras,
  arXiv:2303.17415

\bibitem[{Xu {et~al.}(2018)Xu, Ritson, Ranjan, Todd, Sasselov, \&
  Sutherland}]{C8CC01499J}
Xu, J., Ritson, D.~J., Ranjan, S., {et~al.} 2018, Chem. Commun., 54, 5566

\bibitem[{{Yan} {et~al.}(2021){Yan}, {He}, {Li}, {Esamdin}, {Tan}, {Zhang}, \&
  {Wang}}]{2021MNRAS.505L..79Y}
{Yan}, Y., {He}, H., {Li}, C., {et~al.} 2021, \mnras, 505, L79

\bibitem[{{Yang} \& {Liu}(2019)}]{2019ApJS..241...29Y}
{Yang}, H., \& {Liu}, J. 2019, \apjs, 241, 29

\bibitem[{{Yang} {et~al.}(2017){Yang}, {Liu}, {Gao}, {Fang}, {Guo}, {Zhang},
  {Hou}, {Wang}, \& {Cao}}]{2017ApJ...849...36Y}
{Yang}, H., {Liu}, J., {Gao}, Q., {et~al.} 2017, \apj, 849, 36

\bibitem[{{Yuan} {et~al.}(2023){Yuan}, {Li}, {Bai}, {Dong}, {Wang}, {Yu},
  {Chen}, {Zhao}, {Chu}, \& {Zhang}}]{2023AJ....165..119Y}
{Yuan}, H., {Li}, Z., {Bai}, Z., {et~al.} 2023, \aj, 165, 119

\bibitem[{{Zhao} {et~al.}(2018){Zhao}, {Fan}, {Ren}, {Ge}, {Zhang}, {Li},
  {Wang}, {Wang}, {Qiu}, \& {Jiang}}]{2018RAA....18..110Z}
{Zhao}, Y., {Fan}, Z., {Ren}, J.-J., {et~al.} 2018, Research in Astronomy and
  Astrophysics, 18, 110

\end{thebibliography}
\bibliographystyle{aasjournal}


 \end{CJK*}
\end{document}